\documentclass[prc,twocolumn,epsfig,nofootinbib,floatfix,showpacs]{revtex4}
\usepackage{graphics}
\usepackage{epsfig}
\usepackage{amsfonts}
\usepackage{amsmath}
\usepackage{bm}

\usepackage{color}

\begin{document}
\title{Elimination of spurious modes within QRPA}
\author{A. Repko$^1$, J. Kvasil$^2$, V.O. Nesterenko$^{3,4,5}$}
\affiliation{$^1$
Institute of Physics, Slovak Academy of Sciences, 84511, Bratislava, Slovakia}
\email{anton@a-repko.sk}
\affiliation{$^2$
Institute of Particle and Nuclear Physics, Charles
University, CZ-18000, Praha 8, Czech Republic}
\email{kvasil@ipnp.troja.mff.cuni.cz}
\affiliation{$^3$
Laboratory of Theoretical Physics,
Joint Institute for Nuclear Research, Dubna, Moscow region, 141980, Russia}
\affiliation{$^4$
State University "Dubna", Dubna, Moscow Region, 141980, Russia}
\affiliation{$^5$
Moscow Institute of Physics and Technology,
Dolgoprudny, Moscow region, 141701, Russia}
\email{nester@theor.jinr.ru}

\date{\today}

\begin{abstract}
We suggest a generalized method for elimination of spurious admixtures (SA)
from intrinsic nuclear excitations described within the
Quasiparticle-Random-Phase-Approximation (QRPA). Various kinds of SA-corrections
are treated at the same theoretical ground. The known corrections are well
reproduced. As relevant cases, we consider
subtraction  of SA related with i) violation of
the translational invariance (isovector $E1$ and isoscalar toroidal and compression
$E1$ modes), ii) pairing-induced non-conservation of the particle number
$(E2(K=0)$ and $E0$ modes), and iii) rotational invariance $(E2(K=1)$ and $M1(K=1)$ modes).
The SA subtraction can be done at the level of QRPA states, electromagnetic responses, and
even transition operators.  The additional deformation-induced
corrections for $E1$ excitations are proposed and shown to be essential for
the compression isoscalar mode. The accuracy of the method is demonstrated by Skyrme QRPA
calculations for axially-deformed $^{154}$Sm.

\end{abstract}


\maketitle

\section{Introduction and motivation}

Theoretical analysis of intrinsic nuclear excitations is often complicated
by presence of spurious modes \cite{Th61,TV62,Mar69,Row70,Ri80,BR86}.
These modes appear in the
intrinsic spectra if some symmetries (translational, rotational) and corresponding
conservation laws (for the total momentum  ${\bf P}$ and angular momentum ${\bf J}$)
are violated by the intrinsic Hamiltonian. The spurious modes of this kind represent the motion
of the whole nucleus (translation, rotation) in the laboratory frame. They are obviously beyond
the intrinsic nuclear dynamics and so have to be removed from the intrinsic spectra.
Another particular case is the pairing-induced non-conservation of the proton and
neutron numbers, when the nuclear wave function is contaminated by admixtures
from neighbouring nuclei.

Usually the intrinsic Hamiltonian breaks the conservation laws
in its mean field and pairing parts.
There are many ways for subtraction of emergent spurious admixtures (SA).
For example, in  $E1$ isovector responses, SA are removed using
the proper effective charges  \cite{Ri80}. In the second-order
E1 toroidal and compression isoscalar responses, SA are eliminated
by corrections in the transition operators, requiring the nuclear
center-of-mass to be in rest \cite{VG81,Ha01,Ba94,Kv11}.  There is
also a diversity of projection techniques to exclude SA, e.g.
\cite{Ri80,Be03,Ni06,RRR18,Co00,Ju08,Na07,Mi12,Ars10}.

Random Phase Approximation (RPA) and its quasiparticle version (QRPA) are now
widely used for the self-consistent description of giant resonances and low-energy
states in spherical and deformed nuclei, see e.g.
\cite{Kv11,Be03,Co00,Ju08,Ars10,Na07,Mi12,Re92,Vre00PLB,Vre02tor,Pa07,Kl08,Kv16E0,Te10,Ne16E2}.
As was shown by Thouless \cite{Th61}, RPA has
a principle ability to separate exactly spurious and physical states. In this
method, the spurious mode appears as RPA eigenstate with  zero energy,
which guaranties orthogonality of the spurious and physical
states. Various aspects of this RPA (QRPA) feature are described in detail
elsewhere, see e.g. \cite{TV62,Mar69,Row70,Ri80,Be03,BR86,KvNa86,Na17}.

This RPA advantage to exclude SA was used in early studies within schematic models where
violated symmetries were restored by the proper choice of the residual interaction
compensating the contamination of the mean field \cite{Pya77}. However this scheme
is not relevant for the modern self-consistent RPA methods (Skyrme, Gogny, relativistic)
where the RPA residual interaction is already fully determined by the initial density
functional. Another technique, also not self-consistent,
offers additional terms in the RPA residual interaction to shift the spurious modes outside
the energy region of interest \cite{Pa67,Do05}.

In fully self-consistent RPA (QRPA) models with a complete configuration space,
the spurious modes have to be entirely located at zero-energy eigenstates.
However, even in modern self-consistent RPA calculations this is usually not the case,
see e.g.  discussion in \cite{Be03,Pa07}. Due to a limited size
of configuration space and numerical inaccuracy, we are almost never able to put
the spurious state exactly to the zero energy: its energy is usually a
small positive value. Then the spurious mode is not orthogonal to physical
states and contaminates them, at least neighboring ones. So the problem of
SA persists even in modern self-consistent models.

In the present paper, we propose a simple general method for elimination
of SA from QRPA states and electromagnetic responses. The method
allows to extract arbitrary (related to different symmetries) SA in the framework of
one and the same scheme. We partly use the idea \cite{Ju08,Co00,Na07,Mi12} to
refine QRPA physical
states requiring their orthogonality to the spurious mode. However, in our study,
this idea is realized in a general way using basic QRPA properties. Our scheme
allows to remove  SA not only from QRPA wave functions but also directly from
electromagnetic responses, e.g. by correcting transition operators.
Our amendments to $E1$ responses reproduce well known effective charges and
corrections obtained by various prescriptions \cite{Ri80,Ha01}.
Moreover, we derive and test for these responses the
additional deformation-induced SA-corrections.

In general, the method can be applied to both spherical and deformed nuclei.
We present the formalism and numerical illustrations obtained within the self-consistent
Skyrme QRPA. The main relevant cases are considered: violation of the translational and rotational
invariance and pairing-induced non-conservation of the particle number.
New analytical deformation-induced corrections for SA-elimination from E1 excitations
are derived and tested.

The paper is organized as follows. In Section II, the QRPA background and
detailed description of the method are presented. In Section III,
we demonstrate and discuss examples of SA elimination from $E1$, $E0$,
$E2$, and $M1$ excitations in the axially deformed nucleus $^{154}$Sm.
In Sec. IV, the conclusions are done.

\section{Theoretical framework}

\subsection{QRPA equations}
\label{subsection:A}

In this subsection, we sketch the basics of QRPA formalism \cite{Ri80}
used below in derivation and analysis. We consider
even-even axially-deformed nuclei with the states characterized
by quantum numbers $K^{\pi}$, where $K$ is the component of the
angular momentum to the symmetry axis $z$ and $\pi$ is the space parity.
For nuclear interaction of multipolarity $\lambda\mu$, we have
$K=\mu \ge$0 with $\pi=(-1)^{\lambda}$ for electric
and $\pi=(-1)^{\lambda+1}$ for magnetic modes.

The intrinsic body-fixed Hamiltonian reads
\begin{equation} \label{Hintr}
\hat{H}_{\rm intr} = \hat{H}_{\rm BCS} + \hat{V}_{\rm res}
\end{equation}
where $\hat{H}_{\rm BCS}$ describes mean field and pairing,
$\hat{V}_{\rm res}$ is the residual interaction.

One-phonon QRPA eigenstates $Q^{\dagger}_{\nu}|0 \rangle$
(with $|0 \rangle$ being the QRPA vacuum) are described by the
phonon creation operator
\begin{equation}
\hat{Q}^{\dagger}_{\nu}= \sum_{i>j}
\;
\big( X^{(\nu)}_{ij} \alpha^+_i \alpha^+_j
- Y^{(\nu)}_{ij} \alpha_{\bar{j}}^{\phantom{*}} \alpha_{\bar{i}}^{\phantom{*}} \big)
\label{Q}
\end{equation}
defined as a superposition of $\mathcal{N}$  two-quasiparticle (2qp)
$ij$-excitations with quantum numbers $\mu^{\pi} $. The pairs $ij$ with
$K_i + K_j =\mu$ and $i\bar{j}$ with  $K_i - K_j =\mu$ are used.
The condition $i > j$ means that we involve configurations with
$K_i \ge K_j > 0$;
$\nu$ numerates the phonons with given $\mu^{\pi}$; $X^{(\nu)}_{ij}$ and
$Y^{(\nu)}_{ij}$ are forward and backward 2qp amplitudes;
$|\bar{i}\rangle = \mathcal{T}|i\rangle$
are time-reversed states. The time-reversed counterpart of (\ref{Q}) reads
\begin{equation}
\hat{Q}^{\dagger}_{\bar{\nu}}=\: \sum_{i>j}
\big( X^{(\nu)*}_{ij} \alpha^+_{\bar{i}} \alpha^+_{\bar{j}}
- Y^{(\nu)*}_{ij} \alpha_{j} \alpha_{i} \big) \: .
\label{Qbar}
\end{equation}
The phonon operators obey the features
\begin{equation}
\hat{J}_3 \hat{Q}^{\dagger}_{\nu} |0\rangle = \mu \hat{Q}^{\dagger}_{\nu} |0\rangle , \;
\hat{J}_3 Q^{\dagger}_{\bar{\nu}} |0\rangle = - \mu \hat{Q}^{\dagger}_{\bar{\nu}}|0\rangle
\end{equation}
where $\hat{J}_3$ is z-component of the total momentum $\hat{\bf{J}}$.

Amplitudes $X^{(\nu)}_{ij}$ and $Y^{(\nu)}_{ij}$ and phonon energies
$\hbar \omega_{\nu}$ are obtained from QRPA equations of motion:
\begin{subequations}
\label{RPAeq}
\begin{align}
[\,\hat{H}_{\rm intr},\: \hat{Q}^\dagger_{\nu} \,] &= \hbar \omega_{\nu} \: \hat{Q}^\dagger_{\nu} \: ,
\label{RPAeq_a} \\
[\,\hat{H}_{\rm intr},\: \hat{Q}_{\nu} \,] &= -\hbar \omega_{\nu} \: \hat{Q}_{\nu} \: ,
\label{RPAeq_b} \\
[\, \hat{Q}_{\nu},\: \hat{Q}^\dagger_{\nu^{'}} \, ] & = \delta_{\nu \nu^{'}} .
\label{RPAeq_c}
\end{align}
\end{subequations}
In the matrix form, these equations read  \cite{Ri80}
\begin{equation}
\label{RPAeq_mtrx}
\begin{pmatrix}
A & B  \\
B & A
\end{pmatrix}
\begin{pmatrix}
X^{(\nu)} \\
Y^{(\nu)}
\end{pmatrix}
=
\hbar \omega_{\nu} \:
\begin{pmatrix}
X^{(\nu)} \\
-Y^{(\nu)}
\end{pmatrix} .
\end{equation}
They include real matrices $A$ and $B$:
\begin{subequations}
\label{RPA_AB}
\begin{align}
A_{ij\,i^{'} \! j^{'}} &\equiv (E_{i}+E_j) \, \delta_{ij,\,i^{'} \! j^{'}}
\nonumber \\
& \quad {}+
\langle \rm{BCS}| \, \big[ \alpha_j \, \alpha_i,\:
[ \hat{V}_{res},\, \alpha^+_{i^{'}} \alpha^+_{j^{'}} ] \big] \, |\rm{BCS} \rangle \: ,
\\
B_{ij\:i^{'} \! j^{'}} &\equiv  - \langle \rm{BCS}|
 \, \big[ \alpha_j \, \alpha_i,\, [ \hat{V}_{res},\,
 \alpha_{\bar{j}^{'}} \alpha_{\bar{i}^{'}} ] \big] \, |\rm{BCS} \rangle
\end{align}
\end{subequations}
(where  $|\rm{BCS} \rangle$ is BCS vacuum)  and one-column matrices
\begin{equation}
X^{(\nu)} \equiv
\begin{pmatrix}
\vdots \\
X^{(\nu)}_{ij} \\
\vdots
\end{pmatrix}
\qquad
Y^{(\nu)} \equiv
\begin{pmatrix}
\vdots \\
Y^{(\nu)}_{ij} \\
\vdots
\end{pmatrix}
\quad
ij=1,\cdots ,\mathcal{N} \: .
\label{XY_col}
\end{equation}

According to our time-reversal convention for one-body operators
$\hat{A}$,
\begin{align}
&\mathcal{T}^{-1}\hat{A}\mathcal{T} = \gamma_{\mathcal{T}}^A\hat{A}^\dagger\quad \Rightarrow \nonumber\\
&\langle i|\hat{A}|j\rangle = \gamma_{\mathcal{T}}^A\langle \bar{j}|\hat{A}|\bar{i}\rangle,\quad
\langle i|\hat{A}|\bar{j}\rangle = -\gamma_{\mathcal{T}}^A\langle j|\hat{A}|\bar{i}\rangle \: ,
\label{Tparity}
\end{align}
we introduce time-even ($\gamma_{\mathcal{T}}^A=+1$)
and time-odd ($\gamma_{\mathcal{T}}^A=-1$) operators. Then,
instead of the phonon creation and annihilation operators,
one may define the generalized time-even coordinate
and time-odd momentum operators
\begin{subequations}
\label{Xstr}
\begin{align}
\hat{\mathcal{X}}_{\nu}=
 \sum_{i>j}
 \mathcal{X}^{(\nu)}_{ij}
(\alpha^+_i \alpha^+_j + \alpha_{\bar{j}} \alpha_{\bar{i}}) \; ,
\\
\label{Pstr}
\hat{\mathcal{P}}_{\nu}=
\sum_{i>j}
\mathcal{P}^{(\nu)}_{ij}( \alpha^+_i \alpha^+_j
- \alpha_{\bar{j}} \alpha_{\bar{i}}) \; .
\end{align}
\end{subequations}
Following (\ref{Tparity}), their time-reversed conjugates are
\begin{subequations}
\begin{align}
&\hat{\mathcal{X}}_{\bar{\nu}}^{\phantom{\dagger}} \equiv \hat{\mathcal{X}}_{\nu}^\dagger, \quad
\hat{\mathcal{P}}_{\bar{\nu}}^{\phantom{\dagger}} \equiv -\hat{\mathcal{P}}_{\nu}^\dagger,\\
&\mathcal{X}^{(\bar{\nu})}_{\bar{i}\bar{j}} = \mathcal{X}^{(\nu)*}_{ij}, \quad
\mathcal{P}^{(\bar{\nu})}_{\bar{i}\bar{j}} = \mathcal{P}^{(\nu)*}_{ij}.
\end{align}
\end{subequations}

Operators $\hat{\mathcal{X}}_{\nu}$ and
$\hat{\mathcal{P}}_{\nu}$ are related to the phonon operators (\ref{Q})-(\ref{Qbar}) as
\begin{subequations}
\label{XP_QQ}
\begin{align}
\hat{\mathcal{X}}_{\nu} &= \sqrt{\frac{\hbar}{M_{\nu}
\omega_{\nu}}} \frac{1}{\sqrt{2}} \: \big( \hat{Q}_{\bar{\nu}} + \hat{Q}^\dagger_{\nu} \big) \:,
\\
\hat{\mathcal{P}}_{\nu} &= \frac{\hbar}{i} \:
\sqrt{\frac{M_{\nu} \omega_{\nu}}{\hbar}} \frac{1}{\sqrt{2}}
\: \big( \hat{Q}_{\bar{\nu}} - \hat{Q}^\dagger_{\nu} \big)
\end{align}
\end{subequations}
and vice versa,
\begin{subequations}
\label{QQ_XP}
\begin{align}
\hat{Q}^\dagger_{\nu} &= \sqrt{\frac{M_{\nu} \omega_{\nu}}{2 \hbar}}
\hat{\mathcal{X}}_{\nu} - \frac{i}{\sqrt{2 \hbar M_{\nu} \omega_{\nu}}}
\hat{\mathcal{P}}_{\nu} \: ,
\label{XP_QQ_c} \\
\hat{Q}_{\bar{\nu}} &= \sqrt{\frac{M_{\nu} \omega_{\nu}}{2 \hbar}}
\hat{\mathcal{X}}_{\nu} + \frac{i}{\sqrt{2 \hbar M_{\nu} \omega_{\nu}}}
\hat{\mathcal{P}}_{\nu} \; .
\end{align}
\end{subequations}
 The orthonormalization condition is
\begin{equation}
\big[\, \hat{\mathcal{X}}_{\nu},\: \hat{\mathcal{P}}_{\nu'}^\dagger \, \big] =
- 2 \: \sum_{i>j}
\mathcal{X}^{(\nu)}_{ij} \mathcal{P}^{(\nu')*}_{ij}  =
i \hbar \delta_{\nu \nu^{'}} \; .
\label{XPcomm}
\end{equation}
If $\mathcal{X}^{(\nu)}_{ij}$ is real, then $\mathcal{P}^{(\nu)}_{ij}$
is imaginary, and vice versa. Following (\ref{XP_QQ})-(\ref{XPcomm}),
operators $\hat{\mathcal{X}}_{\nu}$ and
$\hat{\mathcal{P}}_{\nu}$ are defined up to an arbitrary
factor $M_{\nu}$ which cannot be fixed by the normalization condition.

The QRPA equations (\ref{RPAeq}) can be expressed in terms of
$\hat{\mathcal{X}}_{\nu}$ and $\hat{\mathcal{P}}_{\nu}$ as
\begin{subequations}
\label{HXPcomm}
\begin{align}
\big[\,\hat{H}_{\rm{intr}},\, \hat{\mathcal{P}}_{\nu} \,\big] &=
i \hbar M_{\nu}\omega^2_{\nu} \, \hat{\mathcal{X}}_{\nu} \; ,
\label{HXPcomm_a} \\
\big[\,\hat{H}_{\rm{intr}},\, \hat{\mathcal{X}}_{\nu} \,\big] &=
-\frac{i \hbar}{M_{\nu}} \: \hat{\mathcal{P}}_{\nu} \; ,
\label{HXPcomm_b}
\\
\big[\, \hat{\mathcal{X}}_{\nu},\, \hat{\mathcal{P}}_{\nu'}^\dagger \, \big]  &=
i \hbar \: \delta_{\nu \nu^{'}}
\end{align}
\end{subequations}
or, in the matrix form, as
\begin{subequations}
\label{XPmtrx}
\begin{align}
\begin{pmatrix}
A & B  \\
B & A
\end{pmatrix}
\begin{pmatrix}
\mathcal{P}^{(\nu)} \\
\mathcal{P}^{(\nu)}
\end{pmatrix}
&=
i \hbar M_{\nu}  \omega^2_{\nu} \:
\begin{pmatrix}
\mathcal{X}^{(\nu)} \\
\mathcal{X}^{(\nu)}
\end{pmatrix} \; ,
\label{XPmtrx_a}
\\
\begin{pmatrix}
A & B \\
B & A
\end{pmatrix}
\begin{pmatrix}
\mathcal{X}^{(\nu)} \\
- \mathcal{X}^{(\nu)}
\end{pmatrix}
&=
\frac{\hbar}{i} \frac{1}{M_{\nu}} \:
\begin{pmatrix}
\mathcal{P}^{(\nu)} \\
-\mathcal{P}^{(\nu)}
\end{pmatrix}
\; ,
\label{XPmtrx_b}
\end{align}
\end{subequations}
where, in analogy with (\ref{XY_col}),
 $\mathcal{X}^{(\nu)}$ and $\mathcal{P}^{(\nu)}$
are one-column matrices for 2qp amplitudes of the generalized
coordinate and momentum.
For given $K^{\pi}$, the intrinsic Hamiltonian (\ref{Hintr})
in terms $\hat{\mathcal{X}}_{\nu}$ and $\hat{\mathcal{P}}_{\nu}$
has the form
\begin{equation}
\label{HRPA_XP}
\hat{H}_{\rm intr} \approx \hat{H}_{\rm QRPA} =
\sum_{\nu} \!\bigg(\frac{\hat{\mathcal{P}}_{\nu}
\hat{\mathcal{P}}_{\nu}^\dagger}{2 M_{\nu}}
+ \frac{1}{2} M_{\nu}  \omega^2_{\nu}\,
\hat{\mathcal{X}}_{\nu}\hat{\mathcal{X}}_{\nu}^\dagger \bigg) \; .
\end{equation}
This expression shows that the parameter $M_{\nu}$
can be treated as an inertia (mass) value for each QRPA state.

Note that $\langle \hat{\mathcal{X}}, \hat{\mathcal{P}}\rangle$-presentation can
be fruitful for construction of modern self-consistent QRPA versions based
on the functionals with time-even and time-odd densities and currents \cite{Re92},
see particular cases of Skyrme QRPA in \cite{Ne02,Ne06}.

\subsection{Extraction of spurious admixtures}
\label{subsection:B}

The invariance of the Hamiltonian $\hat{H}_{\rm intr}$ under
translation or rotation of the whole nucleus leads
to the conservation condition
\begin{equation}
\big[ \, \hat{H}_{\rm intr},\, \hat{P} \,\big] = 0
\label{commHP}
\end{equation}
where $\hat{P}$ is the corresponding time-odd transformation generator.
The motion of the whole nucleus contaminates the intrinsic nuclear excitations
and leads to SA which have to be extracted from the
intrinsic spectra. For the translation, generator $\hat{P}$ is the
linear momentum operator of the whole nucleus. In axial deformed nuclei,
the center-of-mass translation contaminates the intrinsic $K^{\pi}=0^-$ and
$1^-$ states. For the rotation, $\hat{P}$ is the total angular
momentum operator. The rotation pollutes intrinsic $K^{\pi}= 1^+$ states.

Following \cite{Th61,TV62}, just
$\langle \hat{\mathcal{X}}, \hat{\mathcal{P}}\rangle$-presentation
of QRPA is suitable for the treatment of spurious modes. Indeed,
it is easy to see that the first equation from QRPA set (\ref{HXPcomm})
reproduces condition (\ref{commHP}) if $\omega_0 =0$ and
generator $\hat{P}$  is identified with QRPA generalized momentum
$\hat{\mathcal{P}}_{0}$. Then the Hamiltonian (\ref{HRPA_XP}) recasts to
\begin{equation}
\hat{H}_{\rm QRPA} =
\frac{\hat{\mathcal{P}}_{0}\hat{\mathcal{P}}_{0}^\dagger}{2 M_{0}} +
\sum_{\nu > 0} \!\bigg(\frac{\hat{\mathcal{P}}_{\nu}\hat{\mathcal{P}}_{\nu}^\dagger}{2 M_{\nu}}
 + \frac{1}{2} M_{\nu} \omega^2_{\nu}\,
 \hat{\mathcal{X}}_{\nu}\hat{\mathcal{X}}_{\nu}^\dagger \bigg)
\label{H_XPdec}
\end{equation}
where the spurious  $\nu=0$ state with $\omega_{0} =0$ yields  the
first term \cite{Ri80}. The generalized coordinate $\hat{\mathcal{X}}_{0}$
is obtained from Eq. (\ref{HXPcomm_b}) for given $\hat{\mathcal{P}}_{0}$. Note that
$\hat{\mathcal{X}}_{0}$ is absent in (\ref{H_XPdec}).  The spaces
$\{\hat{\mathcal{X}}_{0}, \hat{\mathcal{P}}_{0},
\hat{\mathcal{X}}_{\nu >0}, \hat{\mathcal{P}}_{\nu >0}\}$ and
$\{ \hat{\mathcal{X}}_{0}, \hat{\mathcal{P}}_{0},
\hat{Q}^{\dagger}_{\nu >0}, \hat{Q}_{\nu >0}\}$ constitute the complete sets of
QRPA states. The condition $\omega_0 =0$ obviously hampers the construction
of well-normalized spurious (s) phonon operator
\begin{equation}
\hat{Q}^{\dagger}_{s} \equiv x_{s} \: \hat{\mathcal{X}}_0 -
\frac{i}{2  \hbar \: x_{s}} \: \hat{\mathcal{P}}_0
\label{Qsp}
\end{equation}
with $x_{s} = \sqrt{M_0 \: \omega_{0}/(2 \hbar)}$. So the {\it exact}
spurious eigenstate can be defined in terms of $\hat{\mathcal{X}}_{0}$  and
$\hat{\mathcal{P}}_{0}$ but not in the phonon representation.

In addition to (\ref{commHP}), one may also consider the conservation law
for the particle number,
\begin{equation}
\label{commHN}
[\,\hat{H}_{\rm intr},\,\hat{N}_{q}\,] = 0 \; ,
\end{equation}
where $\hat{N}_q$ is the time-even particle-number operator for protons
($q = p$)  or  neutrons ($q = n$).
Violation of this law results in spurious admixtures in $K^{\pi}=0^+$ states.
The condition (\ref{commHN}) is held by the QRPA equation (\ref{HXPcomm_b})
if a) $\hat{N}$  is identified with the QRPA generalized coordinate
$\hat{\mathcal{X}}_{0}$ and b) we apply  $M_0 \to {+}\infty$ and
$\omega_0\to 0$ while keeping  $M_0\omega_0^2$ finite (though this
can be hardly realized in practice).

As mentioned in the Introduction, in self-consistent QRPA calculations we are almost never able
to put the energy of the first eigenstate precisely to zero. Even very large 2qp basis
is usually not enough to get $\omega_0=0$. As a result, the conservation laws
(\ref{commHP}) and (\ref{commHN}) are not held precisely and the spurious mode, though
being mainly concentrated in the lowest $\nu =0$ state, still contaminates the neighbouring
physical states with $\nu  > 0$. In other words, the exact spurious mode and
states with $\nu > 0$ are not orthogonal.

Let's suppose that we have almost pure spurious state $|\nu'=0\rangle$  whose energy
$\delta\omega_0$ is yet not zero but a tiny positive value. This state is a reasonable
approximation to the exact spurious mode, i.e. $|\nu'=0\rangle \approx | s \rangle$.
What is important for our aims, the state $|\nu'=0\rangle$ with $\delta\omega_0 > 0$
may be normalized,
\begin{equation}
\big[\, \hat{\mathcal{X}}_{0},\: \hat{\mathcal{P}}_{0}^\dagger \, \big] =
i \hbar  \; ,
\label{XPcomm0}
\end{equation}
and presented in  the phonon form (\ref{Qsp}).

Let's further suppose that we have QRPA self-consistent states
$|\nu \rangle$
contaminated by SA. Our goal is to refine these states from SA.
This can be done requiring orthogonality of the refined states
$|\nu' >0\rangle$ to the spurious mode $| s \rangle$ approximated
by (\ref{Qsp}):
\begin{equation}
\label{ort_nusp}
\langle \nu' | s  \rangle = \langle 0 |\,
\big[\hat{Q}_{\nu'}, \hat{Q}^{\dagger}_{s}\big]\,|0 \rangle = 0
\end{equation}
The phonon operator for $|\nu' \rangle = \hat{Q}^{\dagger}_{\nu'}|0\rangle$
is searched in the form
\begin{equation}
\label{Qab}
\hat{Q}^{\dagger}_{\nu'} = \hat{Q}^\dagger_{\nu} -
\alpha_{\nu} \hat{\mathcal{P}}_0 - \beta_{\nu} \hat{\mathcal{X}}_0
\end{equation}
where $\alpha_{\nu}$ and $\beta_{\nu}$ should be defined from the condition
(\ref{ort_nusp}). This prescription reminds the projection methods used
in some previous studies  for particular spurious
modes, see e.g. \cite{Co00,Ju08,Ars10,Na07,Mi12}. However, as compared with
\cite{Co00,Ju08,Ars10}, we use a more general expression  (\ref{Qsp}) for the spurious state
where both $\hat{\mathcal{X}}_{0}$ and $\hat{\mathcal{P}}_{0}$ operators are included.
As shown below, our way allows to derive a more general scheme for SA-elimination.
There are also essential differences from  \cite{Na07,Mi12}, see discussion
in Sec. IV.D.

The condition (\ref{ort_nusp}) gives
\begin{equation}
\alpha_{\nu} = \bigg[
\frac{\langle 0 |\,\big[ \hat{Q}_{\nu},\, \hat{\mathcal{X}}_0\big]\,|0 \rangle}
{\langle 0 |\,\big[ \hat{\mathcal{P}}^\dagger_0,\,\hat{\mathcal{X}}_0\big]\,|0 \rangle}
\bigg]^*
\; ,
\beta_{\nu} = \bigg[
\frac{\langle 0 |\,\big[ \hat{Q}_{\nu},\, \hat{\mathcal{P}}_0\big]\,|0 \rangle}{\langle 0 |\,\big[ \hat{\mathcal{X}}^\dagger_0,\, \hat{\mathcal{P}}_0\big]\,|0 \rangle}
\bigg]^*
\label{alphabeta}
\end{equation}
or, using (\ref{XPcomm0}),
\begin{equation}
\alpha_{\nu} = \frac{1}{i\hbar}
\langle 0 |\,\big[ \hat{Q}_{\nu},\, \hat{\mathcal{X}}_0\big]\,|0 \rangle^*  ,
\;
\beta_{\nu} = \frac{i}{\hbar}
\langle 0 |\,\big[ \hat{Q}_{\nu},\, \hat{\mathcal{P}}_0\big]\,|0 \rangle^*  .
\label{alphabeta0}
\end{equation}
It is easy to check that
\begin{equation}
\label{Qnucor_proof}
\langle 0 |\,\big[ \hat{Q}_{\nu'},\, \hat{\mathcal{X}}_0\big]\,|0 \rangle =
\langle 0 |\,\big[ \hat{Q}_{\nu'},\, \hat{\mathcal{P}}_0 \big]\,|0 \rangle = 0
\end{equation}
i.e., within the quasiboson approximation, the refined physical states are
indeed orthogonal to the generators $\hat{\mathcal{P}}_0$ and $\hat{\mathcal{X}}_0$
and so  to the spurious state (\ref{Qsp}).
In this derivation, we use the feature that average commutator
of operators with the definite time parity,
\begin{equation}
\label{[A,B]}
\langle 0 |\, [ \hat{A},\,\hat{B} ] \,| 0 \rangle
\propto
 (1 - \gamma_{\mathcal{T}}^A \gamma_{\mathcal{T}}^B) \; ,
\end{equation}
vanishes if operators $\hat{A}$ and $\hat{B}$ have the same time parity
$(\gamma_{\mathcal{T}}^A = \gamma_{\mathcal{T}}^B$)
in the sense (\ref{Tparity}).

Note that result (\ref{Qnucor_proof})
does not depend on the concrete values of $x_{s}$, $M_{\nu}$ and $\omega_0$.
For determination of $\alpha_{\nu}$ and $\beta_{\nu}$,
we should know the symmetry operator and its
conjugate, i.e. $\hat{\mathcal{P}}_{0}$ and $\hat{\mathcal{X}}_{0}$.
These operators are given below in Sec. III for all the cases of interest.

The above scheme allows to refine QRPA states. However in practice we often
need a direct refinement of the transition matrix elements and responses.
This can be done within our approach as well. Let's consider the
matrix element of the transition operator $\hat{\mathcal{M}}$  between the
physical refined state $|\nu' \rangle$ and RPA vacuum:
\begin{align}
\label{Tme_cor}
\langle \nu' |\, \hat{\mathcal{M}} \, | 0 \rangle &
= \langle 0|\, [\hat{Q}_{\nu'},\hat{\mathcal{M}}] \, | 0 \rangle
= \langle 0 |\, \big[ \hat{Q}_{\nu},\, \hat{\mathcal{M}} \big]\, | 0 \rangle
\\
&
-  \alpha_\nu^* \langle 0|\, \big[ \hat{\mathcal{P}}^\dagger_0,\, \hat{\mathcal{M}} \big] \,| 0 \rangle
- \beta_\nu^*  \langle 0|\, \big[ \hat{\mathcal{X}}^\dagger_0,\, \hat{\mathcal{M}} \big] \,| 0 \rangle\,.
\nonumber
\end{align}
Using the feature (\ref{[A,B]}) it is easy to see that, depending on the
time parity of $\hat{\mathcal{M}}$,  the second (third) term in (\ref{Tme_cor}) vanishes at
$\gamma_{\mathcal{T}}^{\mathcal{M}} =-1$ ( $\gamma_{\mathcal{T}}^{\mathcal{M}} =1$). Then
we get
\begin{equation}
\label{Teven_cor}
\langle \nu' |\hat{\mathcal{M}}|0 \rangle =
\langle \nu |\hat{\mathcal{M}}|0 \rangle
-\frac{i}{\hbar}\langle \nu |\hat{\mathcal{X}}_0|0 \rangle
\langle 0 |[\hat{\mathcal{P}}_0^{\dagger}, \hat{\mathcal{M}}]|0 \rangle
\end{equation}
for time-even $\hat{\mathcal{M}}$ and
\begin{equation}
\label{Todd_cor}
\langle \nu' |\hat{\mathcal{M}}|0 \rangle = \langle \nu |\hat{\mathcal{M}}|0 \rangle
+\frac{i}{\hbar}\langle \nu |\hat{\mathcal{P}}_0|0 \rangle
\langle 0 |[\hat{\mathcal{X}}_0^{\dagger}, \hat{\mathcal{M}}]|0 \rangle
\end{equation}
for time-odd $\hat{\mathcal{M}}$.  Expressions (\ref{Teven_cor})-(\ref{Todd_cor})
can be used for calculation of the refined transition matrix elements.

Using (\ref{Teven_cor})-(\ref{Todd_cor}), the refined transition densities
and  currents read:
\begin{eqnarray}
\label{corTD}
\delta\rho_{\nu'}(\bf{r})&=&\delta\rho_{\nu}({\bf r})
-\frac{i}{\hbar}\langle 0 |[Q_{\nu},\hat{\mathcal{X}}_0^{\dagger}]| 0 \rangle
\langle 0 |[\hat{\mathcal{P}}_0^{\dagger}, \hat{\rho}({\bf r})]|0 \rangle ,
\\
\label{corTC}
\delta {\bf j}_{\nu'}({\bf r})&=&\delta {\bf j}_{\nu}({\bf r})
+ \frac{i}{\hbar}\langle 0 |[Q_{\nu},\hat{\mathcal{P}}_0^{\dagger}]| 0 \rangle
\langle 0 |[\hat{\mathcal{X}}_0^{\dagger}, \hat{\bf{j}}({\bf r})]|0 \rangle ,
\end{eqnarray}
where the density and current operators are defined in Appendix A.

One may go further and reduce SA-elimination to modification
of transition operators. Indeed expressions
(\ref{Teven_cor})-(\ref{Todd_cor}) can be rewritten as
$\langle \nu' |\hat{\mathcal{M}}|0 \rangle = \langle \nu |\hat{\tilde{\mathcal{M}}}|0 \rangle$
with
\begin{eqnarray}
\label{Topeven_cor}
\hat{\tilde{\mathcal{M}}} &=& \hat{\mathcal{M}}
-\frac{i}{\hbar} \langle 0 |[\hat{\mathcal{P}}_0^{\dagger}, \hat{\mathcal{M}}]|0 \rangle
\hat{\mathcal{X}}_0 \quad \text{for} \;\gamma_{\mathcal{T}}^{\mathcal{M}}=1 ,
\\
\hat{\tilde{\mathcal{M}}}  &=& \hat{\mathcal{M}}
+\frac{i}{\hbar} \langle 0 |[\hat{\mathcal{X}}_0^{\dagger}, \hat{\mathcal{M}}]|0 \rangle
\hat{\mathcal{P}}_0 \quad \text{for} \;\gamma_{\mathcal{T}}^{\mathcal{M}}=-1 .
\label{Topodd_cor}
\end{eqnarray}

Expressions (\ref{Topeven_cor})-(\ref{Topodd_cor}) suggest the simplest way
for elimination of SA from the responses. They lead to the
important conclusion that QRPA in principle allows to refine
responses through modification  of transition operators.
As compared  with building of the refined QRPA states (\ref{Qab}),  expressions
(\ref{Teven_cor})-(\ref{Topodd_cor})
suggest more economical elimination prescriptions with usage of
initial QRPA states $|\nu \rangle$.

Note that there is an alternative way to obtain expressions
(\ref{Topeven_cor})-(\ref{Topodd_cor}). Since physical and spurious QRPA solutions
form the complete basis, any operator linear in the boson approximation can
be expressed as \cite{Mar69,Row70,Ser03}
\begin{eqnarray}
\hat{\mathcal{M}}&=& \sum_{\nu >0}
\bigg(
\langle 0 |[\hat{\mathcal{Q}}_{\nu}, \hat{\mathcal{M}}]|0 \rangle
\hat{\mathcal{Q}}^{\dagger}_{\nu}
- \langle 0 |[\hat{\mathcal{Q}}_{\nu}^{\dagger}, \hat{\mathcal{M}}]|0 \rangle
\hat{\mathcal{Q}}_{\nu}
\bigg)
\nonumber
\\
&+&\frac{i}{\hbar}
\bigg(
\langle 0 |[\hat{\mathcal{P}}_0^{\dagger}, \hat{\mathcal{M}}]|0 \rangle
\hat{\mathcal{X}}_0
-  \langle 0 |[\hat{\mathcal{X}}_0^{\dagger}, \hat{\mathcal{M}}]|0 \rangle
\hat{\mathcal{P}}_0 \bigg)
\label{Mexpansion}
\end{eqnarray}
where last two terms are spurious contributions. Removal of these
contribution just gives (\ref{Topeven_cor})-(\ref{Topodd_cor}). This
correspondence can be treated as the additional check of the validity of
our projection procedure (\ref{ort_nusp})-(\ref{Qab}). Note that our procedure is more comprehensive
than direct usage of (\ref{Mexpansion}) since it allows to refine not only
operators and their matrix elements but also QRPA wave functions.

Expressions (\ref{Teven_cor})--(\ref{Topodd_cor}) do not include the factor $x_{s}$.
However they need the knowledge of the spurious operators $\hat{\mathcal{P}}_0$ and
$\hat{\mathcal{X}}_0$. As shown in Sec. 4, in some cases, e.g. for $E1$ excitations,
both the symmetry operator and its conjugate are known, and SA corrections acquire
a simple analytical form.  If not, then 2qp amplitudes of the unknown conjugate
and the parameter $M_{0}$ can be determined from equations \cite{Ri80}
\begin{align}
\mathcal{X}^{(0)}_{ij} &= \frac{\hbar}{i} \frac{1}{M_{0}}
\sum_{kl} \big( A - B \big)^{-1}_{ij,\,kl} \, \mathcal{P}^{(0)}_{kl} ,
\label{X_invP}
\\
M_{0} &= 2 \!\sum_{i>j,\,k>l}
\mathcal{P}^{(0)*}_{ij} \, \big( A - B \big)^{-1}_{ij,\, kl} \, \mathcal{P}^{(0)}_{kl}
\label{M_invP}
\end{align}
or
\begin{align}
 \mathcal{P}^{(0)}_{ij} &= i \hbar \omega_{\nu}^2 M_{0}
 \sum_{kl} \big( A + B \big)^{-1}_{ij,\,kl} \, \mathcal{X}^{(0)}_{kl} ,
\label{P_invX}
\\
M_{0} &= \frac{1}{2 \omega_{\nu}^2}\, \bigg\{ \sum_{i>j,\,k>l}
\mathcal{X}^{(0)}_{ij} \big( A + B \big)^{-1}_{ij,\:kl} \,\mathcal{X}^{(0)*}_{kl}\: \bigg\}^{-1} .
\label{M_invX}
\end{align}
As shown in Appendix B, the averages $\langle 0 |[....]|0 \rangle$
in (\ref{Teven_cor})-(\ref{Topodd_cor})
are directly calculated through $\mathcal{P}^{(0)}_{ij}$, $\mathcal{X}^{(0)}_{ij}$
and transition matrix elements.

\section{Details of calculations}

The calculations for axially deformed $^{154}$Sm are performed within
QRPA approach with the Skyrme forces \cite{Re92,Be03,Rep17EPJA}. The total functional
includes the Skyrme, Coulomb and pairing terms. Our approach is fully
self-consistent since both mean field and residual interaction are derived
from the initial functional, using all the available densities and currents.
The Coulomb contribution includes direct and exchange terms in Slater approximation.
The volume pairing is treated at the BCS level. Both particle-hole and pairing-induced
particle-particle channels are involved \cite{Rep17EPJA}. More detail on the
approach are given in the Appendix C. Implementation of the approach to the code
is described in \cite{Re15th,Re15ar}.

We use Skyrme parameterization SLy6 \cite{Cha98} which  was found
successful in our previous QRPA calculations for various dipole  excitations
\cite{Kv11,Kl08,Ne02,Ne06,Kv13,Rep17EPJA,NePRL18,Ne18ar}.
Hartree-Fock (HF) mean field is computed using 2D grid in cylindrical
coordinates (with mesh size 0.4 fm and calculation box of about three nuclear
radii). The single-particle space embraces all the levels from the bottom of the
potential well up to 40 MeV (1533 proton and 1722 neutron levels in $^{154}$Sm).
The volume pairing is treated at the BCS level. The equilibrium quadrupole
deformation $\beta = 0.339$ is obtained by minimization of the system energy.
QRPA calculations  use a large two-quasiparticle (2qp) basis. For example,
for $K^{\pi} = 1^-$ states, the basis includes
$\approx$ 9000 proton and $\approx$ 16000 neutron quasiparticle pairs.
The Thomas-Reiche-Kuhn sum rule \cite{Ri80} and isoscalar dipole energy-weighted
sum rule \cite{Ha01} are exhausted by 95$\%$ and 97$\%$,
respectively.
\begin{figure}[t]
\includegraphics[height=7cm,width=10cm]{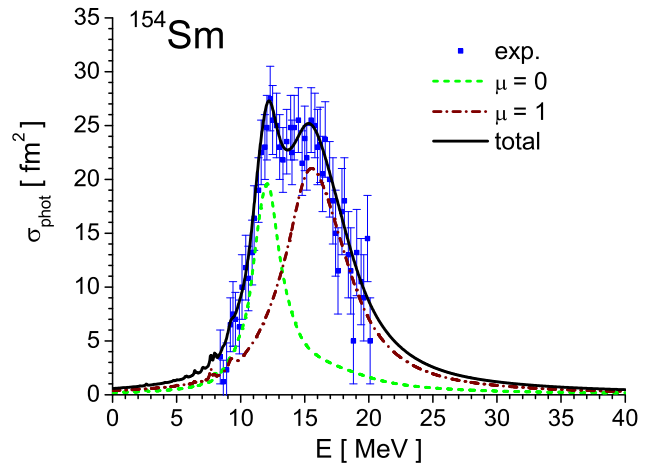}
\caption{ \label{fig:fig1} The total QRPA photoabsorption cross
section (black solid curve) and its $\mu=0$ and $\mu=1$ branches (green dashed
and black dashed-dotted curves) as compared with the experimental data \cite{Gu81}.
}
\end{figure}

Strength function for $X\lambda\mu$-transitions between
the ground state $|0 \rangle$  and  QRPA states $|\nu\rangle$ reads
\begin{equation} \label{sf}
S_k(X\lambda\mu; \, E)
= \sum_{\nu} ( \hbar \omega_{\nu} )^k  \big|\,\langle \nu |
\hat{\mathcal{M}}_{X\lambda\mu}|0 \rangle \, \big|^2 \,
\delta_{\Delta}(E - \hbar \omega_{\nu})
\end{equation}
where $X = E,M$ marks electric and magnetic cases,  $\hbar \omega_{\nu}$ is the excitation
energy of $\nu$-state,  $\langle \nu| \, \hat{\mathcal{M}}_{X \lambda \mu} |0 \rangle$
is the transition matrix element. Components $\mu \ne $ 0 embrace
both $+\mu$ and $-\mu$ contributions.
Further
\begin{equation}
\delta_{\Delta}(E- \hbar \omega_{\nu}) =
\frac{1}{2 \pi} \: \frac{\Delta(\hbar \omega_{\nu})}
{(E- \hbar \omega_{\nu})^2 + [ \Delta(\hbar \omega_{\nu})/2]^2}
\label{Lorentz}
\end{equation}
is the Lorentz weight simulating smoothing effects beyond QRPA
(escape width and coupling to complex configurations). To simulate a general
growth of the smoothing with the excitation energy, the energy-dependent
folding parameter is used \cite{Kv13}:
\begin{equation}
\Delta(\hbar \omega_{\nu}) =
\begin{cases}
\Delta_0 & \text{for } \hbar \omega_{\nu} \leq E_0 \\[2pt]
\Delta_0 + a \, (\hbar \omega_{\nu}-E_0) & \text{for } \hbar \omega_{\nu} > E_0 ,
\end{cases}
\label{deltaEdep}
\end{equation}
where the values $\Delta_0$, $a$, and $E_0$ are adjusted to describe the
experimental photoabsorption cross section. For $^{154}$Sm, these values are
$\Delta_0 = 0.1$ MeV, $a=0.15$, and $E_0=8.3$ MeV.  Then
the calculated photoabsorption cross section
\begin{equation}
\sigma_{\rm phot}(E) = \frac{16 \pi^3}{137\cdot 9\, e^2} \sum_{\mu=0,\pm 1} S_1(E1 \mu;\,E)
\label{sigma_phot}
\end{equation}
well reproduces the experimental data  \cite{Gu81}, see Fig. 1. Here
 the ordinary effective charges $e^{\rm eff}_p=N/A$ and
 $e^{\rm eff}_p=-Z/A$ are used. In the next section, these
charges are derived in the framework of our SA-elimination method.

Being comfortable for description of the high-energy photoabsorption,
the particular energy-dependent folding
(\ref{Lorentz}) with the small low-energy averaging  $\Delta_0 = 0.1$ MeV
is generally not convenient for illustration of
SA-elimination in various low-energy excitations. So, in the next section,
we use the folding with the larger constant averaging $\Delta_0 = 0.4$ MeV.

\section{Results and Discussion}

In this section, we apply our SA-elimination method
to the following particular cases: 1) violation of
the translational invariance ($K^{\pi} = 0^-$ and $1^-$ states;
ordinary,  toroidal (tor) and compression (com) $E1K$ transitions),
2) pairing induced non-conservation of the particle number ($K^{\pi} = 0^+$
states; $E0$ and $E20$ transitions),
3) violation of the rotational invariance  ($K^{\pi} = 1^+$ states;
$E21$ and $M11$ transitions).

We show in subsection \ref{subsec:A} that elimination of SA from ordinary,
compression and toroidal $E1$ responses is reduced  to simple corrections
in the transition operators. Just this simplest
way  is used in numerical calculations. In the cases of non-conservation
of the particle number (subsection \ref{subsec:B}) and violation of the rotational
invariance (subsection \ref{subsec:C}), the building of simple SA-corrections is
hampered. So, in these cases, the numerical results are obtained by formation of
the refined QRPA states (\ref{Qab}). Namely, using the known
symmetry operator and relations (\ref{X_invP})-(\ref{M_invX}),
the set of $\mathcal{P}^{(0)}_{ij}$ and $\mathcal{X}^{(0)}_{ij}$  is obtained.
Then, following (\ref{Tme}), the averages
$\langle 0 |\, \big[ Q_{\nu},\, \hat{\mathcal{M}} \big] \,|0 \rangle$
are calculated
and  coefficients  $\alpha_{\nu}$ and $\beta_{\nu}$ are determined to
construct finally the refined state.

\subsection{$E1$ transitions}
\label{subsec:A}

\subsubsection{Ordinary E1 transitions in the long-wave limit}
\label{subsec:1}

The center-of-mass (CoM) translation of the whole nucleus can
lead to SA in intrinsic dipole nuclear states $K^{\pi} = 0^-$ and $1^-$ .
For $\mu=K$, the operator of CoM linear momentum is
\begin{equation}
\hat{\mathcal{P}}_{0[\mu]} = -i \hbar \sum_{k=1}^{A} (\nabla_{\mu})_k
\label{cmPop}
\end{equation}
where $\mu = 0, \pm 1$,  $\nabla_0 = \frac{\partial}{\partial z}$ ,
$\nabla_{\pm 1} = \mp \frac{1}{\sqrt{2}} \:(\frac{\partial}{\partial x}
\pm i \frac{\partial}{\partial y})$,  $\nabla^\dagger_{\mu} = (-1)^{\mu+1} \nabla_{-\mu}$.
Then, subject to the normalization condition
$[\, \hat{\mathcal{X}}_{0[\mu]} ,\, \hat{\mathcal{P}}_{0[\mu]}^\dagger \,] = i \hbar$,
the CoM coordinate operator has the form
\begin{equation}
\hat{\mathcal{X}}_{0[\mu]} = \sqrt{\frac{4 \pi}{3}} \frac{1}{A}
\sum_{k=1}^{A} (r Y_{1 \mu}(\hat{r}))_k \; .
\label{cmXop}
\end{equation}
The operators (\ref{cmPop}) and (\ref{cmXop}) can be obviously treated as
QRPA operators constituting the spurious state (\ref{Qsp}).
This state has the inertia parameter
$M_0 = mA$ and fully exhausts the energy-weighted sum rule
$EWSR =\frac{3 \hbar^2}{8 \pi m}A$
for isoscalar long-wave dipole excitations.
\begin{figure*}[t]
\vspace{0.4cm}
\includegraphics[height=7cm,width=11cm]{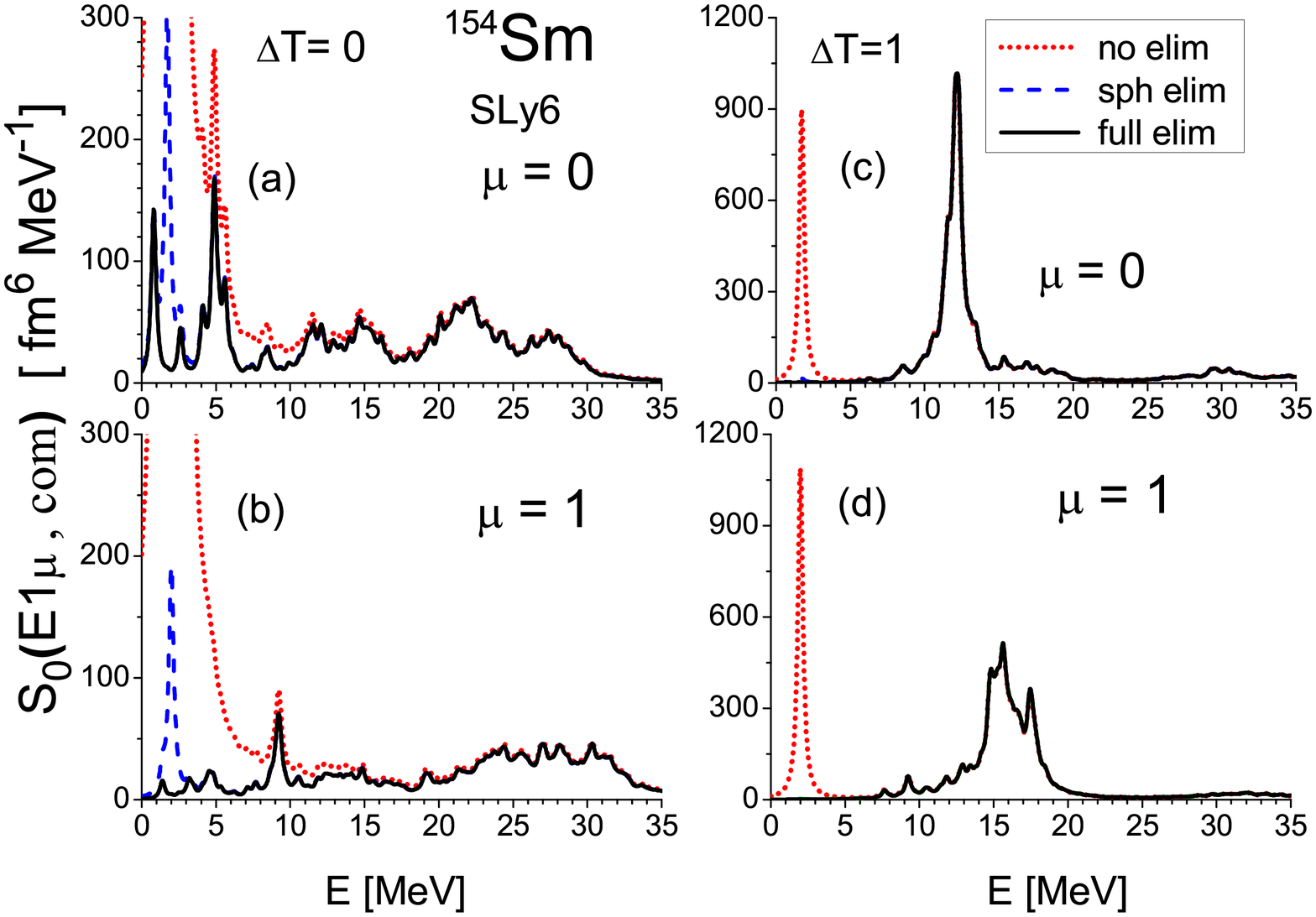}
\caption{ \label{fig:fig2}
The QRPA compression $E1$ strength function (\ref{sf}) in $^{154}$Sm calculated
with the parametrization SLy6. The isoscalar (left panels) and isovector
(right panels) strengths are considered. The branches $\mu=0$ (top), $\mu=1$ (bottom)
are plotted for the polluted strength "no elim" (red dotted curves) and
refined strengths "sph elim" and  "full elim", calculated without
(blue dash curve) and with (black solid curves) deformation correction $d^{\mu}_q$,
respectively.}
\end{figure*}

Now let's consider the proton transition dipole operator in the long-wave limit:
\begin{equation}
\hat{\mathcal{M}}_{E1\mu} = \sum_{k=1}^{Z} (r Y_{1\mu})_k .
\end{equation}
Following (\ref{Topeven_cor}), this time-even operator is refined as
\begin{equation}
\label{T_cmc}
\hat{\tilde{\mathcal{M}}}_{E1\mu} = \hat{\mathcal{M}}_{E1\mu}
-\frac{i}{\hbar}
\langle 0 |[\hat{P}^{\dagger}_{0[\mu]}, \hat{\mathcal{M}}_{E1\mu}]|0 \rangle
\hat{X}_{0[\mu]}.
\end{equation}
Using the relation
$[\hat{P}^{\dagger}_{0[\mu]}, \hat{\mathcal{M}}_{E1\mu}]
= - i \hbar Z \sqrt{\frac{3}{4\pi}}$
and Eq. (\ref{cmXop}) for $\hat{X}_{0[\mu]}$,
the above expression is reduced to
\begin{eqnarray}
\hat{\tilde{\mathcal{M}}}_{E1\mu} &=&
\sum_{k=1}^{Z} (r Y_{1\mu})_k
-\frac{Z}{A}\sum_{k=1}^{A} (r Y_{1\mu})_k
\nonumber
\\
&=&
\frac{N}{A}\sum_{k=1}^{Z} (r Y_{1\mu})_k
-\frac{Z}{A}\sum_{k=1}^{N} (r Y_{1\mu})_k.
\label{ME1corr}
\end{eqnarray}
So, for $E1\mu$-transitions, we get the standard
effective charges  $e^{\rm eff}_p=N/A$ and $e_n^{\rm eff}=-Z/A$,  This
justifies validity of our method in this particular case.
The dipole strength function obtained with this effective charges
is demonstrated in Fig. 1.

\subsubsection{Compression E1 transitions}
\label{subsec:2}

The transition operator for $E1$ compression  mode (CM) is \cite{Ha01}
\begin{equation}
\hat{\mathcal{M}}_{E1\mu , \rm{com}}^{(\Delta T)}=
\frac{1}{10} \sum_{q=n,p} e^{(\Delta T)}_q \sum_{k \in q}
(r^3 Y_{1\mu})_k \; .
\label{T_E1com}
\end{equation}
The effective charges are
\begin{eqnarray}
\label{eeff}
e^{(\Delta T)}_p &=& e^{(\Delta T)}_n = 1 \quad\text{for isoscalar case} \quad \Delta T = 0,
\\
e^{(\Delta T)}_p &=& -e^{(\Delta T)}_n = 1 \quad \text{for isovector case} \quad \Delta T = 1.
\nonumber
\end{eqnarray}
Usually CM is observed in the isoscalar reaction $(\alpha, \alpha')$  \cite{Ha01},
so the channel $\Delta T$=0 is most relevant. However, for the completeness, we also consider
the case  $\Delta T$=1. Operator (\ref{T_E1com}) originates from  the second-order term in the long-wave
decomposition of the total electric $E1$ transition operator, see \cite{Kv11} for more detail.
$E1$ compression mode can be affected by CoM motion. The spurious QRPA momentum
(\ref{cmPop}) and coordinate  (\ref{cmXop}) operators are obviously the same as for ordinary
$E1$ transitions considered above.

Since the transition operator (\ref{T_E1com}) is time-even, its refined
version is determined by Eq. (\ref{Topeven_cor}):
\begin{eqnarray}
\label{Tcom_cmc}
\hat{\tilde{\mathcal{M}}}^{(\Delta T)}_{E1\mu, \rm{com}} &=&
\hat{\mathcal{M}}^{(\Delta T)}_{E1\mu, \rm{com}}
\\
&-&\frac{i}{\hbar}
\langle 0 |[\hat{P}^{\dagger}_{0[\mu]}, \hat{\mathcal{M}}^{(\Delta T)}_{E1\mu, \rm{com}}]|0 \rangle
\hat{X}_{0[\mu]}.
\nonumber
\end{eqnarray}
Using relations (\ref{r3Y10_der})-(\ref{Y00})
for vector spherical harmonics, given in Appendix D,
we get
\begin{eqnarray}
\label{PM_com}
&&\langle 0 |[\hat{P}^{\dagger}_{0[\mu]},
\hat{\mathcal{M}}^{(\Delta T)}_{E1\mu, \rm{com}}]|0 \rangle
=-\frac{i\hbar}{10\sqrt{3}}
\\
&&\sum_{q=n,p} e^{(\Delta T)}_q
\{
\frac{5}{2\sqrt{\pi}}\langle r^2\rangle_q
- \frac{2}{\sqrt{5}} c_{\mu}
\langle r^2 Y_{20}\rangle_q \} ,
\nonumber
\end{eqnarray}
where $c_{\mu}$=-2 for $\mu$=0 and 1 for $\mu=\pm 1$. Besides,
\begin{equation}
\label{rav}
  \langle r^2\rangle_q =\int d^3r \rho_q ({\bf r}) r^2, \;
  \langle r^2 Y_{20}\rangle_q =\int d^3r \rho_q ({\bf r}) r^2 Y_{20}
\end{equation}
where $\rho_q ({\bf r})$ is the proton or neutron density.

Substitution of  (\ref{cmXop})  and (\ref{PM_com}) into (\ref{Tcom_cmc})
yields
\begin{eqnarray}
\label{MTcom_cmc}
\hat{\tilde{\mathcal{M}}}^{(\Delta T)}_{E1\mu, \rm{com}}&=&
\hat{\mathcal{M}}^{(\Delta T)}_{E1\mu, \rm{com}}
\\
&-& \frac{1}{10A} \hat{D}_{E1\mu} \sum_{q=n,p} e^{(\Delta T)}_q
\left(\frac{5}{3}\langle r^2\rangle_q - d^{\mu}_q \right)
\nonumber
\end{eqnarray}
where
\begin{eqnarray}
\hat{D}_{E1\mu} &=& \sum_{k=1}^{A} (r Y_{1\mu})_k \, ,
\\
\label{d_def}
d^{\mu}_q &=&
\frac{4}{3}\sqrt{\frac{\pi}{5}} c_{\mu}
\langle r^2 Y_{20}\rangle_q \, .
\end{eqnarray}
The correction $d^{\mu}_q$ with $\langle r^2 Y_{20}\rangle_q$
appears only in nuclei with an axial quadrupole deformation. To our knowledge,
this is the first derivation of the deformation-induced CoM correction for
the dipole compression operator.

In the important isoscalar $\Delta T=0$ case, we have
\begin{align}
&\hat{\tilde{\mathcal{M}}}^{(0)}_{E1\mu, \rm{com}} =
\hat{\mathcal{M}}^{(0)}_{E1\mu, \rm{com}}
 -\frac{1}{10}  \hat{D}_{E1\mu} \left(\frac{5}{3}\langle r^2\rangle_0 - d^{\mu}_0 \right)
 \nonumber
 \\
 &= \frac{1}{10}\sum_{k=1}^{A}
 \bigg[ r^3_k-r_k \left(\frac{5}{3}\langle r^2\rangle_0 - d^{\mu}_0 \right)\bigg]
 Y_{1\mu}(\hat{r}_k)
 \label{T0com_cmc}
\end{align}
where
\begin{equation}
\langle r^2\rangle_0 = \frac{1}{A} (\langle r^2\rangle_p + \langle r^2\rangle_n), \,
d^{\mu}_0 = \frac{1}{A} (d^{\mu}_p + d^{\mu}_n) ,
\end{equation}
and
\begin{equation}
\langle r^2 Y_{20}\rangle_0 =
\frac{1}{A} (\langle r^2 Y_{20}\rangle_p + \langle r^2 Y_{20}\rangle_n)
\approx 5/(4\pi) \beta \langle r^2\rangle_0
\end{equation}
with $\beta$ being the deformation parameter.
The term $\sim \langle r^2\rangle_0$ in (\ref{T0com_cmc}) precisely
reproduces the familiar CoM correction for $E1$ CM operator in spherical nuclei
\cite{VG81,Ha01,Kv11,Ba94,Co00,Vre00PLB,Pa07}. This confirms
the validity of our approach.

In the isovector case $\Delta T=1$, we get
\begin{equation}
\label{T1com_cmc}
\hat{\tilde{\mathcal{M}}}^{(1)}_{E1\mu, \rm{com}} =
\hat{\mathcal{M}}^{(1)}_{E1\mu, \rm{com}}
 -\frac{1}{10}  \hat{D}_{E1\mu} \left(\frac{5}{3}\langle r^2\rangle_1 - d^{\mu}_1 \right)
\end{equation}
with $\langle r^2\rangle_1 = 1/A (\langle r^2\rangle_p-\langle r^2\rangle_n)$
and $d^{\mu}_1 = 1/A (d^{\mu}_p-d^{\mu}_n)$. So the CoM correction persists
in the isovector E1 CM as well.

In Fig.2, we demonstrate elimination of SA from $E1$ compression strength
functions (\ref{sf}) in $^{154}$Sm. The strength function has no any energy
multiplier and so actually represents the reduced transition probability
$B(E1\mu, {\rm com},\Delta T)=|\langle \nu |
\hat{\mathcal{M}}_{E1\mu}^{\Delta T}|0 \rangle \, \big|^2$. We see
$\mu=0$ and $\mu=1$ strengths for $\Delta T=0$ and  $\Delta T=1$ channels,
computed with the transition operators (\ref{T0com_cmc}) and (\ref{T1com_cmc}).
The following cases are shown: "no elim" - without SA-elimination
corrections inside the parentheses in (\ref{T0com_cmc}) and (\ref{T1com_cmc}),
"sph elim"  - using only spherical part  of the
corrections ($d^{\mu}_{0,1}=0$),  "full elim" - using the full corrections.

Figure 2 shows that, for $\Delta T=0$, the SA-pollution is absent
at $E>$ 15 MeV, noticeably changes the strength
at  4 - 8 MeV $< E <$ 15 MeV, and gives a huge spurious peak at 0 $< E <$ 4 MeV.
Both SA-eliminations, spherical and full,
suppress the lowest spurious peak and drastically change the low-energy
CM strength. What is remarkable, the spherical ($d^{\mu}_{0}=0$)
and full ($d^{\mu}_{0} \ne 0$) corrections result in very different
low-energy spectra: concentrated in one peak in "sph elim" and fragmented
in "full elim". So, for low-energy CM$(\Delta T=0)$, the
deformation-induced correction  $d^{\mu}_{0}$ is very important.

Right plots of Fig. 2 demonstrate SA-elimination in isovector CM.
In this case, the spurious mode is concentrated
in one significant peak at a few MeV and is negligible at a higher
energy. In general, the pollution effect for $\Delta T=1$ is much smaller than for
$\Delta T=0$. This is not surprising since the spurious mode is isoscalar
and so should contaminate mainly $\Delta T=0$ strength. We see
that  the low-energy spurious peak is fully removed by
our SA-corrections. The options "sph elim" and "full elim" give
almost indistinguishable strengths, i.e. the impact of $d^{\mu}_{1}$
is negligible.

\subsubsection{Toroidal $E1$ transitions}
\label{subsec:3}

The toroidal E1 transition dipole operator \cite{Kv11} is
\begin{eqnarray}
\hat{\mathcal{M}}_{E1\mu,\, \rm{tor}}^{(\Delta T)}&=&
{-}\frac{1}{2 \sqrt{3}}
 \int d^3r \, r^2
\nonumber \\
 \cdot \Big( \hat{\bf j}^{(\Delta T)}({\bf r})
 & \cdot &  \Big[ {\bf Y}^{0}_{1 \mu}(\hat{r}) +
 \frac{\sqrt{2}}{5} {\bf Y}^{2}_{1 \mu}(\hat{r}) \Big] \Big) \; ,
\label{T_E1tor}
\end{eqnarray}
where ${\bf Y}^{0}_{1 \mu}$ and ${\bf Y}^{2}_{1 \mu}$ are
vector spherical harmonics. Operator of the nuclear current
$\hat{\bf j}^{(\Delta T)}=\hat{\bf j}^{(\Delta T)}_{\rm c} +
\hat{\bf j}^{(\Delta T)}_{\rm m}$  is the sum of the convective
and magnetization parts, see Appendix A.
Effective  charges $e^{(\Delta T)}_q$ are defined in (\ref{eeff}).
The toroidal operator (\ref{T_E1tor}) is just
the second-order term in  the long-wave
decomposition of the total electric $E1$ transition operator \cite{Kv11}.

For $E1$ toroidal mode (TM), the spurious QRPA momentum and coordinate
 operators are again the same as for ordinary and compression $E1$
modes considered above, i.e. are given by Eqs. (\ref{cmPop})-(\ref{cmXop}).
However, unlike the previous  $E1$ cases,
the toroidal $E1$ transtion operator is time-odd in the sense
(\ref{Tparity}). This can be easily recognized
taking into account the time-odd character of the nuclear current (\ref{j_nuc}).
Then, following Eq. (\ref{Topodd_cor}), the refined transition toroidal operator is
\begin{eqnarray}
\label{Ttor_cmc}
\hat{\tilde{\mathcal{M}}}^{(\Delta T)}_{E1\mu, \rm{tor}} &=&
\hat{\mathcal{M}}^{(\Delta T)}_{E1\mu, \rm{tor}}
\\
&+&\frac{i}{\hbar}
\langle 0 |[\hat{X}^{\dagger}_{0[\mu]}, \hat{\mathcal{M}}^{(\Delta T)}_{E1\mu, \rm{tor}}]|0 \rangle
\hat{P}_{0[\mu]}.
\nonumber
\end{eqnarray}
Note that $\hat{\mathcal{M}}^{(\Delta T)}_{E1\mu, \rm{tor}}$ includes
the total nuclear current. At the same time, the magnetization current
$\hat{\bf j}^{(\Delta T)}_{\rm m}$ does not contribute to the commutator average
$\langle 0 |[\hat{X}^{\dagger}_{0[\mu]},
\hat{\mathcal{M}}^{(\Delta T)}_{E1\mu, \rm{tor}}]|0 \rangle$ and so to the SA-correction.

Using  (\ref{Y0Y2})-(\ref{Y0Y0}), the commutator average in
(\ref{Ttor_cmc}) can be written as
\begin{eqnarray}
\label{XM_com}
&&\langle 0 |[\hat{X}^{\dagger}_{0[\mu]},
\hat{\mathcal{M}}^{(\Delta T)}_{E1\mu, \rm{tor}}]|0 \rangle
\\
&=& - i \frac{1}{4\sqrt{3\pi}} \frac{e\hbar}{m} \frac{1}{A} \sum_{q=n,p} e^{(\Delta T)}_q
[ \langle r^2\rangle_q + \frac{3}{10}d^{\mu}_q ]
\nonumber
\end{eqnarray}
where $\langle r^2\rangle_q$ and deformation correction $d^{\mu}_q$ are defined in
(\ref{rav}) and (\ref{d_def}).
Then, using the relation
\begin{equation}
\hat{P}_{0[\mu]}=\sqrt{4\pi}\frac{m}{e}
\int d^3r (\hat{\bf j}^{(0)}_c({\bf r}) \cdot {\bf Y}^0_{1\mu}({\hat r}))
\end{equation}
with the isoscalar convective current  $\hat{\bf j}^{(0)}_c$,
the refined transition toroidal operator (\ref{Ttor_cmc}) acquires
the form
\begin{eqnarray}
\label{Ttor_cmc1}
\hat{\tilde{\mathcal{M}}}^{(\Delta T)}_{E1\mu, \rm{tor}} &=&
-\frac{1}{2\sqrt{3}} \int d^3r \,
\\ \nonumber
\cdot \Big[
\Big( \hat{\bf j}^{(\Delta T)}({\bf r})
&\cdot &
[ \frac{\sqrt{2}}{5} {\bf Y}^{2}_{1 \mu}(\hat{r})
 + {\bf Y}^{0}_{1 \mu}(\hat{r})] r^2
\Big)
 \\
 - \frac{1}{A}
 \Big( \hat{\bf j}^{(0)}_c({\bf r}) &\cdot& {\bf Y}^{0}_{1 \mu}  \Big)
  \sum_{q=n,p} e^{(\Delta T)}_q
(\langle r^2\rangle_q + \frac{3}{10}d^{\mu}_q )
\Big] \; .
\nonumber
\end{eqnarray}
For $\Delta T$=0 transitions and neglecting $\hat{\bf j}^{(0)}_{\rm m}$, we obtain
\begin{align}
\label{Ttor0_cmc1}
& \hat{\tilde{\mathcal{M}}}^{(0)}_{E1\mu, \rm{tor}} =
-\frac{1}{2\sqrt{3}} \int d^3r \, \Big( \hat{\bf j}^{(0)}_c({\bf r})
\cdot [ \frac{\sqrt{2}}{5} r^2 {\bf Y}^{2}_{1 \mu}(\hat{r})
\\
& + {\bf Y}^{0}_{1 \mu}(\hat{r})(r^2 -
\langle r^2\rangle_0 - \frac{3}{10}d^{\mu}_0 ) ]\Big) .
\nonumber
\end{align}
The term $\sim \langle r^2\rangle_0$ in (\ref{Ttor0_cmc1}) precisely
reproduces the ordinary CoM correction for E1 toroidal operator,
obtained earlier for spherical nuclei \cite{Kv11,Vre02tor,Pa07}.
This once more confirms the validity of our approach.
Note that the previous derivation of this correction exploits
some approximate relations following from sum rules, see e.g. \cite{Kv11}.
Instead, the present derivation is free from such approximations.

In the isovector  $\Delta T$=1 case, the refined operator is
\begin{eqnarray}
\label{Ttor1_cmc1}
\hat{\tilde{\mathcal{M}}}^{(1)}_{E1\mu, \rm{tor}} &=&
\hat{\mathcal{M}}^{(1)}_{E1\mu, \rm{tor}}
\\
 + \frac{1}{2\sqrt{3}} &\int d^3r&
 \Big( \hat{\bf j}^{(0)}_c({\bf r}) \cdot {\bf Y}^{0}_{1 \mu}  \Big)
(\langle r^2\rangle_1 + \frac{3}{10}d^{\mu}_1 ) .
\nonumber
\end{eqnarray}
where $\langle r^2\rangle_1$ and  $d^{\mu}_1$ are defined in
the previous subsection for CM. Note that, despite the transition
operator is isovector, its SA-correction is determined by the isoscalar current
operator $\hat{\bf j}^{(0)}_c$.

\begin{figure}[t]
\vspace{0.4cm}
\includegraphics[height=6.5cm,width=8.5cm]{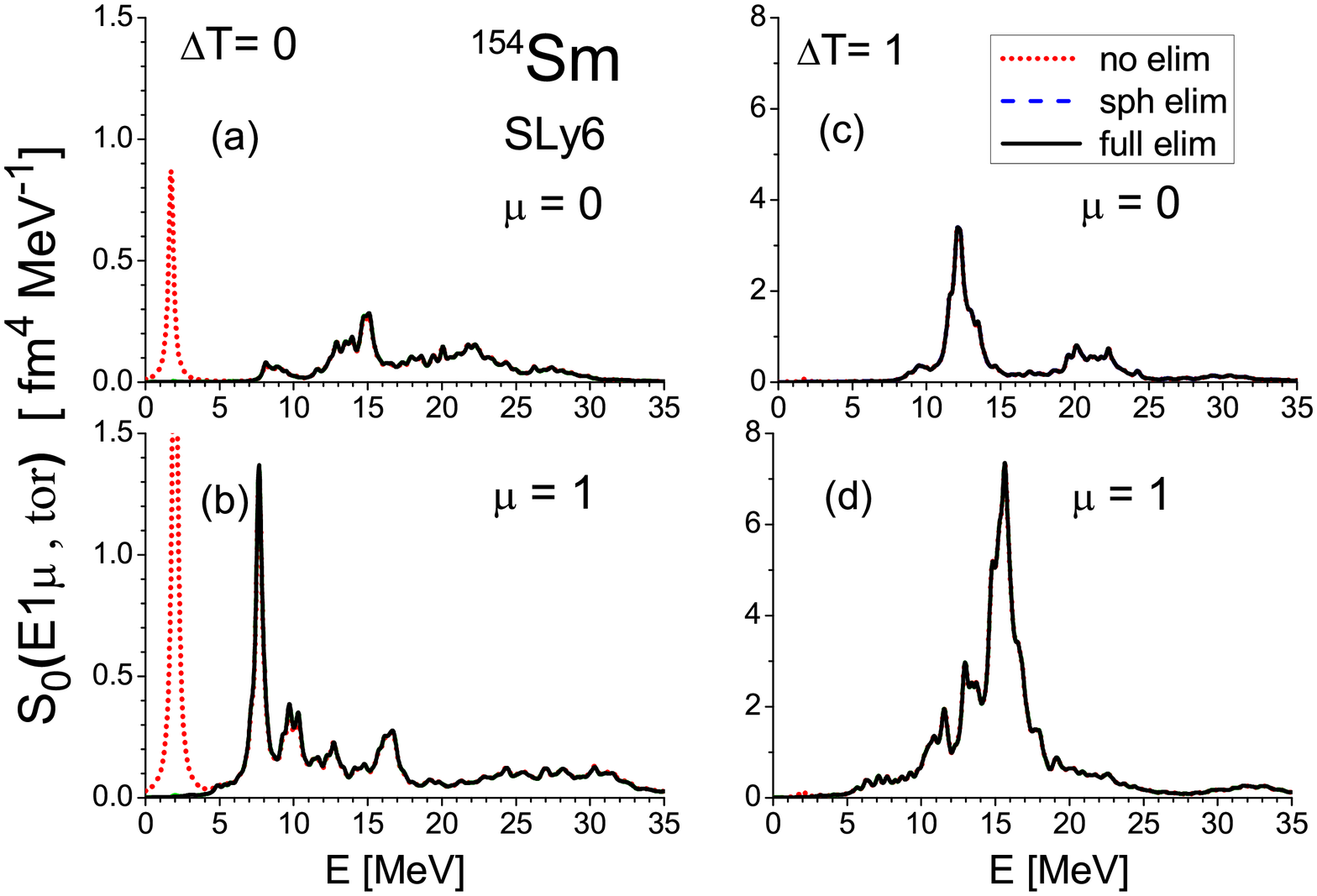}
\caption{ \label{fig:fig3}
The same as in the Fig.2 but for toroidal $E1$ response.
}
\end{figure}

 In Fig. 3, the SA-elimination effect for the toroidal $E1$ excitations is illustrated.
Unlike CM$(\Delta T=0)$ case in Fig. 2, SA in TM($\Delta T=0$) strength
is almost fully concentrated in the lowest peak while the strength at a higher energy
is not contaminated. The difference
in SA-pollution for CM and TM is explained by different character of these modes. CM is
irrotational and so is strongly affected by CoM which is also irrotational. Instead TM is
basically vortical \cite{Kv11} and  so the CoM impact on TM is much less. For
TM($\Delta T=1$), the pollution is almost absent.
Fig. 3 shows that our procedure fully suppresses spurious peaks in TM$(\Delta T=0)$.
In all the plots, the green dashed and black solid lines are practically coincide, i.e.
effect of the deformation-induced corrections $d^{\mu}_{0,1}$  is negligible. This is
explained by a small (as compared to CM) relative weight of $d^{\mu}_{0,1}$
in the toroidal SA-corrections given in (\ref{Ttor0_cmc1})-(\ref{Ttor1_cmc1}).

\subsection{Elimination of SA from $E0$ and $E20$ excitations}
\label{subsec:B}

The pairing treated  within Bardeen–Cooper–Schrieffer (BCS)
procedure leads to violation of the conservation law (\ref{commHN})
for the particle number $N_q$ \cite{Ri80}. This results in spurious
admixtures in electric monopole $E0$ and quadrupole $E20$ excitations
with $K=0$. Time-even  operators for E0 and E20 transitions are
\begin{equation}
\hat{\mathcal{M}}_{E\lambda 0}^{(\Delta T)}=
\sum_{q=n,p} e^{(\Delta T)}_q \sum_{k \in q} (r^2 Y_{\lambda 0})_k
\label{E00E20}
\end{equation}
with $\lambda$=0 and 2.

The symmetry operator is the time-even  operator
of the particle number $\hat{N}_q$. It can be  associated with the spurious
operator $\hat{X}_0$ (for simplicity of notation, we omit below the index $q$).
Then, within BCS, we get
\begin{equation}
  \hat{X}_0 =\hat{N} =  \sum_{j>0} \mathcal{X}^{(0)}_{j{\bar j}}
\big( \alpha^+_j \alpha^+_{\bar{j}} +  \alpha_{\bar{j}} \, \alpha_j \big)
\label{XNpair}
\end{equation}
with $\mathcal{X}^{(0)}_{j\bar{j}}= 2\: \mathcal{U}_j \mathcal{V}_j $ and
$\mathcal{U}_j$, $\mathcal{V}_j$ being BCS pairing amplitudes.
Following (\ref{Pstr}) the time-odd conjugate spurious operator is
\begin{equation}
\hat{\mathcal{P}}_0 = \sum_{ij} \mathcal{P}^{(0)}_{ij} \,
\big( \alpha^+_i \alpha^+_{j} -  \alpha_{\bar{j}} \alpha_{\bar{i}} \big) \, .
\label{PNpair}
\end{equation}
Since the values  $\mathcal{X}^{(0)}_{j{\bar j}}$ are known, we can obtain
their conjugates $\mathcal{P}^{(0)}_{ij}$ from (\ref{P_invX}).
Then, using  (\ref{Tme}) from Appendix B, we can calculate the averages
$\langle 0 |\, \big[ Q_{\nu},\, \hat{\mathcal{M}} \big] \,|0 \rangle$,
determine the coefficients  $\alpha_{\nu}$ and $\beta_{\nu}$,
and finally build refined states.
Just this prescription was used to get the numerical results shown
in this subsection. It partly reminds
the earlier projection scheme  proposed  for E0 excitations in \cite{Ju08}.
However, our prescription is more general. As shown below,
we also suggest the direct refinement of
the transition matrix elements and operators.
\begin{figure}
\vspace{0.3cm}
\includegraphics[height=6.5cm,width=8.5cm]{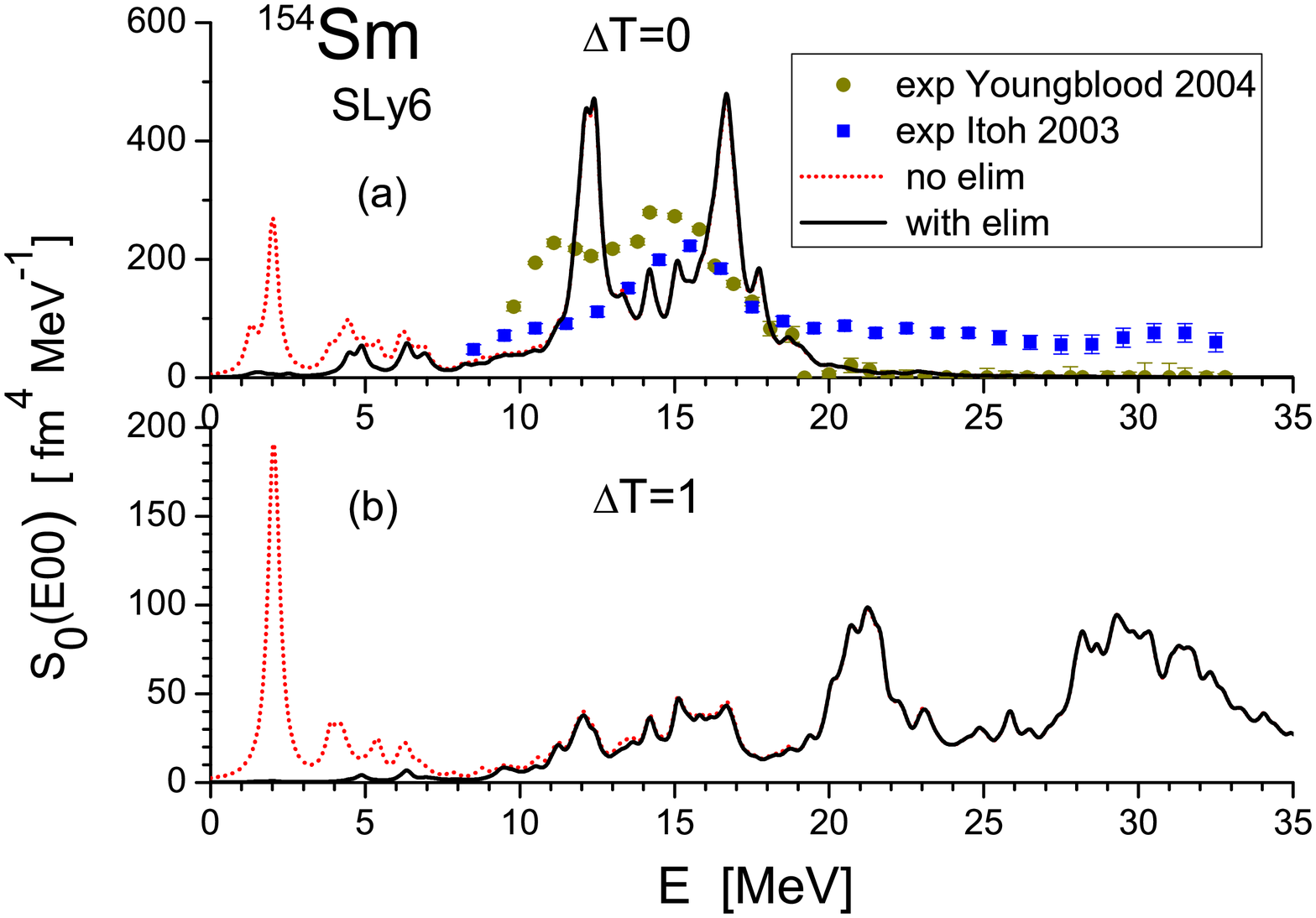}
\vspace{0.2cm}
\caption{ \label{fig:fig4}
Skyrme QRPA strength function (\ref{sf}) for isoscalar (top panel) and
isovector (bottom panel)  $E0$ transitions. Results for $\Delta T$=0
are compared with $(\alpha,\alpha')$ experimental data of D.H.
Youngblood et al \cite{Yo04} and M. Itoh et al \cite{It03}.
The strengths without  (red dotted curve) and with (black solid curves) SA-elimination
are shown.
}
\end{figure}
\begin{figure*}
\includegraphics[height=7cm,width=11cm]{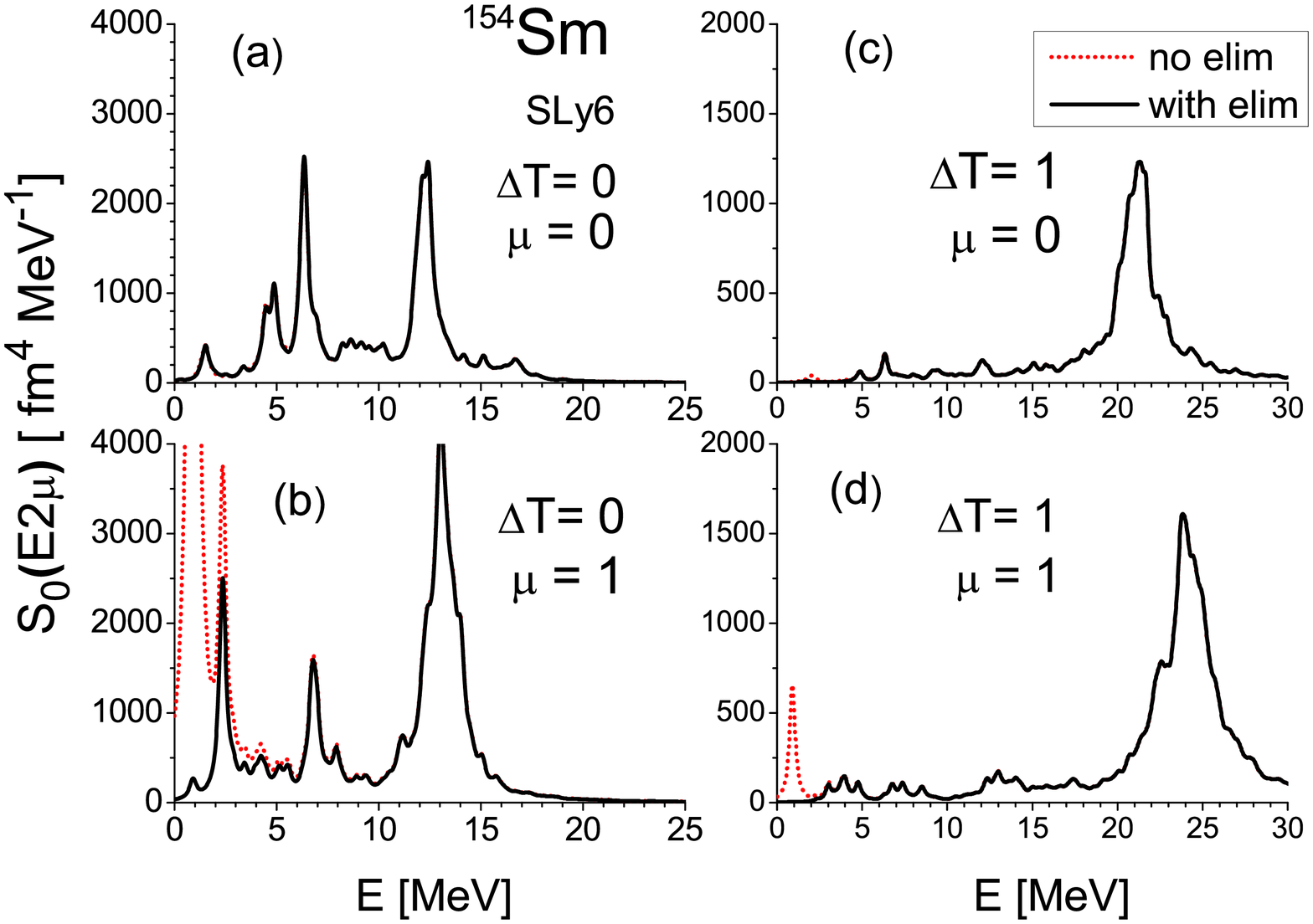}
\caption{ \label{fig:fig5}
Isoscalar (left) and isovector (right) $E20$ ($\mu$=0) and $E21$ ($\mu$=1)
strength functions calculated without (red dotted curves) and with
(black solid curves) SA-elimination.}
\end{figure*}

As an alternative way, we can also  construct
the refined transition operator. Following (\ref{Topeven_cor}), it reads
\begin{eqnarray}
\label{TO_0020}
\hat{\tilde{\mathcal{M}}}^{(\Delta T)}_{E\lambda 0} &=&
\hat{\mathcal{M}}^{(\Delta T)}_{E\lambda 0}
-\frac{i}{\hbar}
\langle 0 |[\hat{P}^{\dagger}_{0}, \hat{\mathcal{M}}^{(\Delta T)}_{E\lambda 0}]|0 \rangle
\hat{N}
\\
&=& \hat{\mathcal{M}}^{(\Delta T)}_{E\lambda 0}
- \gamma^{(\Delta T)}_{E\lambda 0}
\hat{N}.
\nonumber
\end{eqnarray}
with
\begin{eqnarray}
\gamma^{(\Delta T)}_{E\lambda 0} &=&
\frac{i}{\hbar}
\langle 0 |[\hat{P}^{\dagger}_{0}, \hat{\mathcal{M}}^{(\Delta T)}_{E\lambda 0}]|0 \rangle
\\
&=&  \frac{2i}{\hbar} \sum_{ij} [\mathcal{P}^{(0)}_{ij}]^*
\langle ij |\hat{\mathcal{M}}^{(\Delta T)}_{E\lambda 0}|0 \rangle .
\nonumber
\end{eqnarray}
In quasiparticle representation,
\begin{equation}
\hat{\tilde{\mathcal{M}}}^{(\Delta T)}_{E\lambda 0} =
\sum_{ij}
\langle ij| \hat{\tilde{\mathcal{M}}}^{(\Delta T)}_{E\lambda 0}|0\rangle
\big( \alpha^+_i \alpha^+_{j} +  \alpha_{\bar{j}} \, \alpha_{\bar{i}} \big)
\end{equation}
with
\begin{equation}
\langle ij| \hat{\tilde{\mathcal{M}}}^{(\Delta T)}_{E\lambda 0}|0\rangle =
\langle ij| \hat{\mathcal{M}}^{(\Delta T)}_{E\lambda 0}|0\rangle
- 2 \gamma^{(\Delta T)}_{E\lambda 0}
\delta_{\bar{i}j}\mathcal{U}_j \mathcal{V}_j .
\end{equation}
So, to exclude SA, the transition operator can be corrected only
in matrix elements with $|j\bar{j}\rangle$.
The factor $\mathcal{U}_j \mathcal{V}_j$ is large only for
states near the Fermi level and, therefore, just these states
mainly contribute to the correction.

The SA-elimination effect for the monopole strength $S_0(E0)$ and
quadrupole strength $S_0(E20)$ is demonstrated in Figs. 4 and 5.
Both strengths embrace the same set of QRPA $K^{\pi}=0^+$ states
calculated with the particle-particle channel \cite{Rep17EPJA}.

Fig. 4 illustrates elimination of SA from the isoscalar and isovector
$E0$ responses calculated with effective charges from (\ref{eeff}).
In the upper panel, the calculated strength
$S_0(E0)$ rather well reproduces experimental data \cite{Yo04}. Both features,
two-hump structure of the Giant Monopole Resonance (GMR) and vanishing
the strength above GMR, are described. At the same time, $S_0(E0)$ deviates
in these features from  the data  \cite{It03}. There is a definite
discrepancy between the data \cite{Yo04} and \cite{It03}, though
both of them are obtained from $(\alpha,\alpha')$ reaction,
see discussion in \cite{Kv16E0}.

Figure 4 shows that in both $\Delta T =0$ and $\Delta T =1$ channels
SA are not merely concentrated in the lowest peak
but essentially contaminate low-energy excitations at
$E < $8 MeV. At the higher energy,
the pollution is negligible.
Our procedure successfully removes the spurious strength.

In the top panels of Fig. 5, the quadrupole strength $S_0(E20)$ with and
without SA-elimination is demonstrated. In contrast to E0 case, the SA
contamination is almost negligible.
Some elimination effect is visible only for the minor
lowest spurious peak at $E\approx$ 2 MeV in $\Delta T$=1 channel.
The difference in the pollution of $E0$ and $E2$ strengths can be explained
by basically monopole character of the pairing. So just $E0$  but not $E2$
strength is contaminated.

\subsection{Elimination of SA from $E21$ and $M11$ excitations}
\label{subsec:C}

The rotational invariance is related with the conservation of the total
angular momentum $\hat{\bf J}$ of the nucleus \cite{Mar69,Row70,Ri80}.
Since in axial deformed nuclei
the rotation around the intrinsic symmetry z-axis is forbidden, the conservation law
is formulated through $\mu=\pm1$ components of $\hat{\bf J}$, combining
intrinsic x- and y-axes:
\begin{equation}
\big[\, \hat{H}_{\rm{intr}}, \hat{J}_{\mu=\pm1} \,\big] = 0 .
\label{HJcon}
\end{equation}
Below we consider only
\begin{equation}
\hat{J}_1 = -\frac{1}{\sqrt{2}} (\hat{J}_x + i \hat{J}_y) = \hat{L}_1 + \frac{1}{2} \sigma_{1}
\label{68}
\end{equation}
where $\hat{L}_1$ and $\sigma_{1}$ are
components of the angular momentum  and Pauli matrix for $\mu$=1.
The violation of the conservation law (\ref{HJcon}) leads to SA
in $K^{\pi}=1^+$ states and contaminates $E21$ and $M11$ transitions
between these states and the ground state.
The symmetry operator $\hat{J}_1$ is time-odd and so can be associated
with the spurious operator
$\hat{\mathcal{P}}_0$ \cite{Mar69,Ri80,KvNa86}:
\begin{equation}
\hat{J}_{1} = \hat{\mathcal{P}}_0 = \sum_{ij} \mathcal{P}^{(0)}_{ij}
\, \big( \alpha^+_{i} \alpha^+_{j} - \alpha_{\bar{j}} \alpha_{\bar{i}} \big)
\label{69}
\end{equation}
where $\mathcal{P}^{(0)}_{ij} = \langle ij |\, \hat{J}_{1} \,|0 \rangle$
are real two-quasiparticle matrix elements.
The angle operator $\hat{\Theta}$, being the time-even conjugate to $\hat{J}_{1}$,
matches the spurious operator $\hat{\mathcal{X}}_0$ \cite{Ri80}:
\begin{equation}
\hat{\Theta}_1 \equiv \hat{\mathcal{X}}_0 =
\sum_{ij} \mathcal{X}^{(0)}_{ij} \, \big( \alpha^+_{i} \alpha^+_{j}
 + \alpha_{\bar{j}} \alpha_{\bar{i}} \big)
\label{70}
\end{equation}
where $\langle ij |\, \hat{\mathcal{X}}_0 \,|0 \rangle$ are
imaginary. The operators obey the normalization condition
$\big[\, \hat{\mathcal{X}}_0,\,
\hat{\mathcal{P}}_0^\dagger \,\big] =
\big[\, \hat{\Theta}_1,\, \hat{J}_{1}^\dagger\, \big] =i \hbar$.
Using known matrix elements
$\mathcal{P}^{(0)}_{ij} = J^{(1)}_{ij} = \langle ij |\, \hat{J}_{1} \,|0 \rangle$,
the values $\mathcal{X}^{(0)}_{ij}$ are obtained from the inversion equation (\ref{X_invP}).
Then, as in the previous subsection, we can calculate the averages
$\langle 0 |\, \big[ Q_{\nu},\, \hat{\mathcal{M}} \big] \,|0 \rangle$,
determine the coefficients  $\alpha_{\nu}$ and $\beta_{\nu}$,
and finally construct the refined QRPA states. This way was utilized to get
the numerical results shown below.

The parameter $M_0$ is calculated combining (\ref{M_invP}) with
(\ref{XPcomm0}). It  has the physical meaning of the
principal $\mu=1$ component of the moment of inertia $\mathcal{F}_1$
\cite{Mar69,Ri80,KvNa86}:
\begin{equation}
M_0 = \mathcal{F}_1 \approx 2
\sum_{ij,kl}\! J^{(1)\,*}_{ij} \, (A-B)^{-1}_{ij,\,kl} \, J^{(1)}_{kl} \;.
\label{71}
\end{equation}
\begin{figure}
\includegraphics[height=8cm]{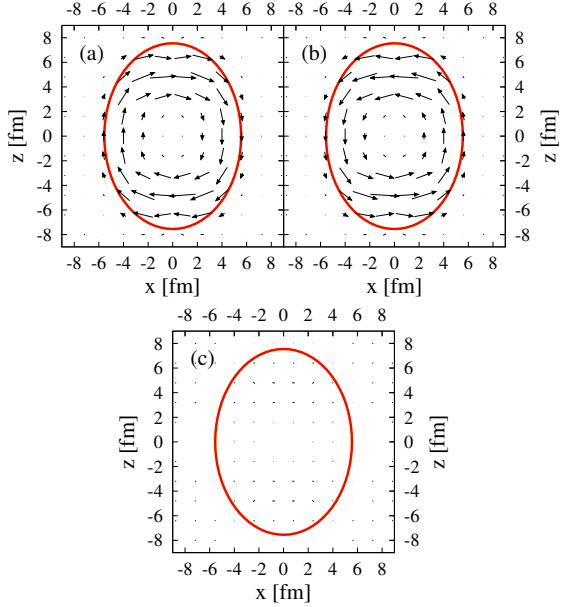}
\caption{
The convective  isoscalar current transition density in the x-z plane
of the intrinsic frame, calculated following (\ref{corTC})
for the first ($\nu=0$) spurious QRPA $K^{\pi}=1^+$ solution
with the energy $\hbar \omega_{0}=\:0.95$ MeV. The panels show:
(a) $\delta {\bf j}_{\nu=0}^{(\Delta T=0)}({\bf r})$  without SA-elimination, (b)
 the SA-correction term in the r.h.s of (\ref{corTC}), (c)
 $\delta {\bf j}_{\nu'=0}^{(\Delta T=0)}({\bf r})$ with SA-elimination.}
\label{fig:fig6}
\end{figure}

The $E21$ and $M11$ transition operators are characterized
by the time-even operator
\begin{equation}
\hat{\mathcal{M}}_{E21}^{(\Delta T)}=
\sum_{q=n,p} e^{(\Delta T)}_q \sum_{k \in q} (r^2 Y_{21})_k
\label{E21}
\end{equation}
and the time-odd operator
\begin{equation}
\hat{\mathcal{M}}_{M11}^{(\Delta T)} =
\frac{e \hbar}{2 m c} \sqrt{\frac{3}{4\pi}}\,
\sum_{q=n,p} \sum_{k \in q}
\big[  e^{(\Delta T)}_q \, \hat{l}^{(k)}_{1} + g_q \, \hat{s}^{(k)}_{1} \big]
\label{M11}
\end{equation}
where $\hat{l}^{(k)}_{1}$  and $\hat{s}^{(k)}_{1}$ are ($\mu$=1)-components
of operators of the orbital momentum and spin for $k$-th nucleon. Further,
$e^{(\Delta T)}_q$ are effective charges. They are taken as (\ref{eeff}) for $E21$
and as $e^{(\Delta T)}_p$=1 and $e^{(\Delta T)}_n$=0  for $M21$.
Gyromagnetic factors $g_q=g^s_q\eta$ are composed from the  nucleon bare
factors $g^s_q$ with the quenching parameter $\eta$=0.7 \cite{Ha01}.

The corresponding refined operators are
\begin{eqnarray}
\label{TE21_cmc}
\hat{\tilde{\mathcal{M}}}^{(\Delta T)}_{E21} &=&
\hat{\mathcal{M}}^{(\Delta T)}_{E21}
-\frac{i}{\hbar}
\langle 0 |[\hat{J}_1^{\dagger}, \hat{\mathcal{M}}^{(\Delta T)}_{E21}]|0 \rangle
\hat{\Theta}_1 \; ,
\\
\hat{\tilde{\mathcal{M}}}^{(\Delta T)}_{M11} &=&
\hat{\mathcal{M}}^{(\Delta T)}_{M11}
+ \frac{i}{\hbar}
\langle 0 |[\hat{\Theta}_1^{\dagger}, \hat{\mathcal{M}}^{(\Delta T)}_{M11}]|0 \rangle
\hat{J}_1 \; .
\label{TM11_cmc}
\end{eqnarray}
Following Appendix B, the average commutators
in (\ref{TE21_cmc}) and  (\ref{TM11_cmc}) can be computed as
\begin{eqnarray}
\langle 0 |[\hat{J}_1^{\dagger}, \hat{\mathcal{M}}^{(\Delta T)}_{E21}]|0 \rangle
 &=& 2\,\sum_{ij} [\mathcal{P}_{ij}^{(0)}]^*\,
 \langle ij | \, \hat{\mathcal{M}}^{(\Delta T)}_{E21} \,|0 \rangle ,
\\
\langle 0 |[\hat{\Theta}_1^{\dagger}, \hat{\mathcal{M}}^{(\Delta T)}_{M11}]|0 \rangle
 &=& 2\,\sum_{ij} [\mathcal{X}_{ij}^{(0)}]^*\,
 \langle ij | \, \hat{\mathcal{M}}^{(\Delta T)}_{M11} \,|0 \rangle .
\end{eqnarray}
SA-corrections in (\ref{TE21_cmc}) and (\ref{TM11_cmc}) include $\hat{J}_1$ and,
in this sense, correspond to the  corrections suggested earlier in
\cite{Mar69,Ri80,KvNa86}.

In the bottom ($\mu=1$) plots of Fig. 5, we demonstrate subtraction of SA from $E21$
responses. The plots show the strong elimination effect for low-energy states,
especially in $\Delta T =0$ channel.

To illustrate the elimination mechanism, we show in Fig. 6 the convective part of
the isoscalar current transition density (\ref{corTC})
for the spurious $K^{\pi}=1^+$ state at 0.95 MeV (this state is depicted
by the dotted red line in $\Delta T=0$ and $\Delta T=1$ bottom plots of Fig. 5).
Following (\ref{corTC}), the refined current transition density
$\delta \bf{j}_{\nu'=0}$ is the sum of the initially
polluted $\delta \bf{j}_{\nu=0}$ and the correction current.
In the plot (a) with $\delta \bf{j}_{\nu=0}$,
we see a clear spurious rotation. The SA-correction current shown in the plot (b)
demonstrates the opposite rotation. These two currents  compensate each other
and thus give the vanishing $\delta \bf{j}_{\nu'=0}$ in the plot (c).
\begin{figure}
\includegraphics[height=6cm,width=8.5cm]{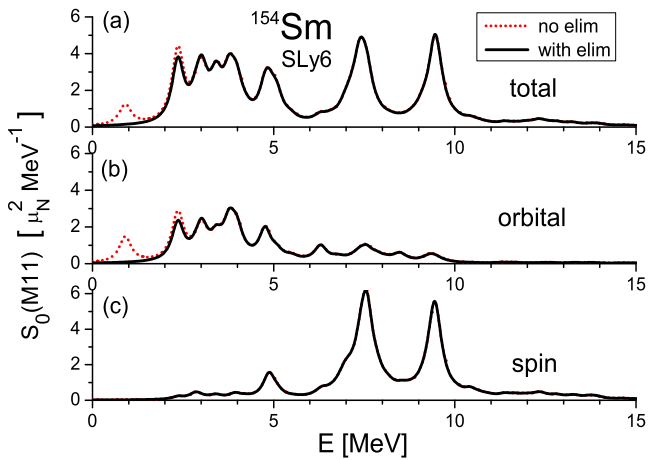}
\caption{ \label{fig:fig7}
The total (top), orbital (middle) and spin (bottom) $M11$ strength
functions calculated without (red dotted line) and with (black
solid line) SA-elimination.}
\end{figure}

Further, Fig. 7 shows the SA-elimination effect for $M1(K=1)$ strength in
$^{154}$Sm. The strength functions are calculated for $M11$ transition
operator (\ref{M11}) consisting of the orbital and spin parts.
As seen from the figure, orbital part of $M11$-operator generates
$M1(K=1)$ orbital scissor mode located at 2-4 MeV (plot (b)) while
spin part of the operator produces the spin-flip resonance lying at
6-11 MeV (plot (c)). We see that the spuriosity caused by
the nuclear rotation concerns only the low-energy orbital part of $M11$
strength. Our SA-elimination method fully suppresses the
spurious peak at 0.95 MeV in the orbital $M11$ strength.

\subsection{Comparison with other approaches}
\label{subsec:D}

In our method, we get the refined QRPA physical states from the condition
(\ref{ort_nusp})-(\ref{Qab}), i.e.  requesting orthogonality of physical states to
the spurious mode (which can be also considered as projection of contaminated states
onto refined physical states). This simple and evident condition was also used in some
previous works, e.g. \cite{Co00,Ju08,Ars10,Na07,Mi12}. Let' briefly compare
our and previous studies.

The works \cite{Co00,Ju08,Ars10} consider subtraction of SA
from compression $E1(\Delta T=0)$ \cite{Co00}, monopole $E0(\Delta T=0)$
 \cite{Ju08}, and dipole
$E1(\Delta T=1)$\cite{Ars10} QRPA states in spherical nuclei. The principle
difference of our approach with these works is that we use a more general
expression (\ref{Qsp}) for the spurious state where both $\hat{\mathcal{X}}_{0}$
and  $\hat{\mathcal{P}}_{0}$ operators are included. This allows us to get,
in the same theoretical frame, the general SA-subtraction recipe
covering various symmetry violations.

In works \cite{Na07,Mi12}, the spurious state embraces both
time-even and time-odd parts. However these works deal with
specific QRPA versions: finite amplitude method \cite{Na07}
and Green’s function method \cite{Mi12}. So their recipes have specific
forms determined by particular QRPA realizations. Besides these recipes address
only transition densities \cite{Na07} or strength functions \cite{Mi12}.
Instead,  our method is based on the conventional matrix  QRPA and
suggests SA-elimination for a wider set of characteristics: wave functions,
transition matrix  elements (transition densities), and  transition operators.

Altogether, the major differences and advantages of our method as compared to
the previous studies \cite{Co00,Ju08,Na07,Mi12,Ars10}, can be summarized as:

a) Unlike \cite{Co00,Ju08,Ars10,Na07,Mi12}, we propose SA-corrections
at different stages of the calculations: for QRPA states, matrix elements and even
transition operators. Various symmetry violations can be covered, both
spherical and deformed  nuclei can be considered. This flexibility is indeed
important in practical calculations.

b) Our method reproduces well known SA-corrections for
conventional $E1(\Delta T=1)$ \cite{Ri80},
compression $E1(\Delta T=0)$ \cite{VG81,Kv11} and toroidal
$E1(\Delta T=0)$ \cite{Vre02tor,Kv11} excitations,
obtained earlier in different models. In \cite{Co00,Ju08,Ars10,Na07,Mi12},
these corrections are considered as independent items to be compared
with the projection results. We show that the previous corrections
can be derived on the same theoretical footing within the
projection technique. This deepens our knowledge on the
nature and accuracy of SA-elimination in dipole states.

c) For the first time, deformation-induced analytical corrections for $E1$ compression
and toroidal transitions were derived and numerically tested.  They were
found to be essential for compression $E1(\Delta T=0)$ low-energy excitations.

\section{Conclusion}

A general simple  method for elimination of spurious admixtures (SA) from RPA/QRPA
intrinsic nuclear excitations is proposed. The SA-corrections are derived from
the requirement of orthogonality of physical QRPA states to the
phonon-like spurious state. Within this projection technique,
the most relevant cases are inspected: violation
of the translational invariance (ordinary, compression and toroidal
$E1$ modes), pairing-induced
non-conservation of the particle number ($E2(K=0)$ and $E0$ modes), and
violation of the rotational invariance ($E2(K=1$) and $M1(K=1)$ modes).
Various familiar SA-corrections are rederived on the same
theoretical footing and new elimination schemes are proposed.

For each relevant case, the SA-subtraction is illustrated by
Skyrme QRPA calculations for axially deformed  $^{154}$Sm. High
efficiency and accuracy of the method are demonstrated.

The method is universal. Both isoscalar ($\Delta T$ = 0) and isovector ($\Delta T$ = 1)
excitations are covered. The refinement from SA can be carried out
at different levels of calculations: for each RPA/QRPA state  and directly
for various electric and magnetic responses.
In the later case, the SA-corrections are derived for transition matrix elements
and even for transition operators. For $E1$ excitations,
the analytical expressions for SA-corrections are proposed.
For axial deformed nuclei, the additional deformation-induced
SA-corrections for the compression and toroidal $E1$ strengths are derived.
It is shown that these corrections  are important for the low-energy part
of the $E1(\Delta T$ = 0) compression mode. The method can be applied to
various RPA/QRPA approaches including self-consistent ones.

\section*{Acknowledgement}

The work was partly supported by Votruba-Blokhincev (Czech Republic-BLTP JINR).
A.R. is grateful for support from Slovak Research and Development
Agency under Contract No. APVV-15-0225. J.K. acknowledges the grant
of Czech Science Agency (project 19-14048S).

\appendix

\section{Operators of nuclear density and current}
\label{app:A}

The density operator is
\begin{equation}
\hat{\rho}^{(\Delta T)}({\bf r}) = e \sum_{q=n,p} e^{(\Delta T)}_q
\sum_{k \in q}\delta({\bf r} - {\bf r}_k)
\label{rho_nuc}
\end{equation}
with the effective charges  $e^{(0)}_p=e^{(0)}_n = 1$ in the isoscalar ($\Delta T$=0) case
and $e^{(1)}_p=-e^{(1)}_n = 1$ in the isovector ($\Delta T$=1) case.

The operator of the nuclear current
\begin{equation}
\hat{\bf j}^{(\Delta T)}({\bf r}) =
\hat{\bf j}_{\rm c}^{(\Delta T)}({\bf r})
+ \hat{\bf j}_{\rm m}^{(\Delta T)}({\bf r})
\label{j_nuc}
\end{equation}
consists from the convective and magnetization parts
\begin{eqnarray}
\hat{\bf j}_{\rm c}^{(\Delta T)} ({\bf r}) &=&
-i\frac{e \hbar}{2 m}\sum_{q=n,p} e^{(\Delta T)}_q
\\
&\cdot&
\sum_{k \in q}
\big[\delta({\bf r} - {\bf r}_k)\, {\bf \nabla}_k
+ {\bf \nabla}_k\,\delta({\bf r} - {\bf r}_k)\big],
\nonumber
\\
\hat{{\bf j}}_{\rm m}^{(\Delta T)}({\bf r}) &=& \frac{e \hbar}{2 m}
 \sum_{q=n,p} g_q \sum_{k \in q}
{\bf \nabla} \times  \hat{{\bf s}}_k \; \delta({\bf r} - {\bf r}_k ) .
\label{38}
\end{eqnarray}
Here  $m$ is the nucleon mass, $\hat{s}$ is the spin operator,
$g_q$ is the nucleon gyromagnetic factor.

\section{Commutator averages}
\label{app:B}

If $\hat{\mathcal{M}} = \sum_{ij}\langle ij |\hat{\mathcal{M}}|{\rm BCS} \rangle
(\alpha^+_i \alpha^+_j +\gamma_{\mathcal{T}}^{\mathcal{M}}  \alpha_{\bar{j}} \alpha_{\bar{i}})$,
then averages in
(\ref{Teven_cor}) -
(\ref{Topodd_cor})
have the form:
\begin{align}\label{Tme}
&\langle 0 |\, \big[ Q_{\nu},\, \hat{\mathcal{M}} \big] \,|0 \rangle =
\\
&\quad=
\sum_{ij} \big[ X^{(\nu)*}_{ij} + \gamma_{\mathcal{T}}^{\mathcal{M}} Y^{(\nu)*}_{ij} \big] \,
\langle ij|\, \hat{\mathcal{M}} \,|{\rm BCS} \rangle \: ,
\nonumber
\end{align}
\begin{align} \label{T_Xpme}
&\langle 0 | \, \big[ \hat{\mathcal{X}}_0^\dagger ,\, \hat{\mathcal{M}} \big] \, |0 \rangle =
\\
&\ =
\begin{cases}
     0  \qquad \qquad \qquad \qquad \qquad \quad \text{ for time-even } \hat{\mathcal{M}}  \\[2pt]
     \begin{array}{l}
       2\,\sum_{ij} \mathcal{X}_{ij}^{(0)\,*}\,\langle ij | \, \hat{\mathcal{M}} \,|{\rm BCS} \rangle \:
       \text{ for time-odd }            \hat{\mathcal{M}} \; ,
  \end{array}
\end{cases}
\nonumber
\end{align}
\begin{align} \label{T_Ppme}
&\langle 0 | \, \big[ \hat{\mathcal{P}}_0^\dagger ,\, \hat{\mathcal{M}} \big] \, | 0 \rangle =
\\
&\ =
   \begin{cases}
     \begin{array}{l}
       2\,\sum_{ij} \mathcal{P}_{ij}^{(0)\,*}\,\langle ij | \, \hat{\mathcal{M}} \,|{\rm BCS} \rangle \:  \text{ for time-even }           \hat{\mathcal{M}}
      \end{array}  \\[2pt]
    0 \qquad \qquad \qquad \qquad \qquad \quad   \text{ for time-odd } \hat{\mathcal{M}} \;,
   \end{cases}
\nonumber
\end{align}
where $|{\rm BCS} \rangle$ is the BCS vacuum.

\section{Skyrme QRPA framework}
\label{app:C}

The total functional ${\mathcal E}_{\rm tot}$ includes Skyrme, Coulomb
and pairing parts  \cite{Re92,Be03,Rep17EPJA}:
\begin{equation}
{\mathcal E}_{\rm tot} =
{\mathcal E}_{\rm Sk} +
{\mathcal E}_{\rm Coul} +
{\mathcal E}_{\rm pair} \; .
\end{equation}
The Skyrme part ${\mathcal E}_{\rm Sk}\{J^{\varsigma}_q\}$
depends on the set  $\{J^{\varsigma}_q\}$ of densities and currents
(listed by $\varsigma$) for protons and neutrons ($q$=p,n). This set
includes time-even (nucleon $\rho_q$,  kinetic-energy $\tau_q$, spin-orbit
${\bf J}_q$) and time-odd (current ${\bf j}_q$, spin ${\bf s}_q$,
vector kinetic-energy ${\bf T}_q$) items.
The Coulomb functional ${\mathcal E}_{\rm Coul}(\rho_p)$  consists from the direct
term and exchange terms in Slater approximation \cite{Be03,Rep17EPJA}.

The pairing functional ${\mathcal E}_{\rm pair}(\rho_q)$
can be taken in the surface and volume forms, i.e. with and without
the density dependence \cite{Be00,Rep17EPJA}.
For simplicity reasons, we consider here the volume form
\begin{equation}
\mathcal{E}_{\rm pair} = \frac{1}{4} \: \sum_{q=n,p} V_q \:
\int d^3r \: |\tilde{\rho}_q ({\bf r})|^2
\label{Epair}
\end{equation}
where $V_q$ are neutron and proton pairing constants and
\begin{equation}
\tilde\rho_q({\bf r}) = 2 \sum_{i \epsilon q} f_i^q v_i u_i |\psi_i ({\bf r})|^2
\label{rho_pair}
\end{equation}
are pairing densities with single-particle wave functions
$\psi_i ({\bf r})$, Bogoliubov pairing  factors $v_i$ and  $u_i$
and energy-dependent cut-off weights $f^q_i$ \cite{Rep17EPJA}.

The nuclear mean field is determined by Hartree-Fock method using
first functional derivatives
$\delta ({\mathcal E}_{\rm Sk}+{\mathcal E}_{\rm Coul})/\delta J^{\varsigma}_q$
over time-even densities $J^{\varsigma}_q$. The volume pairing is treated
within the BCS scheme \cite{Rep17EPJA}.

The residual interaction is determined by the second functional derivatives
$\delta^2 {\mathcal E}_{\rm tot}/\delta J^{\varsigma}_q \delta J^{\varsigma'}_{q'}$
\cite{Rep17EPJA}.
The contributions of all time-even and time-odd densities and currents,
including the pairing density (\ref{rho_pair}), is taken into account. Both
particle-hole (ph) and pairing-induced particle-particle (pp) channels are involved,
see detailed expressions in \cite{Rep17EPJA}. In  ph-channel, Coulomb
contribution is included.

Our QRPA approach is fully self-consistent since i) both the mean field and residual
interaction are obtained from the same initial functional, ii) contributions
of all the densities and currents are taken into account, iii) both ph- and pp-channels
are considered.

\section{Useful relations}
\label{app:D}

In derivation of SA corrections to E1 transition operators,
the following relations were used \cite{Va76}:
\begin{align}
\nabla_0 r^3 Y_{10}(\hat{r}) &= \frac{1}{\sqrt{3}} \,
\Big[ 5 r^2 Y_{00} + \frac{4}{\sqrt{5}} r^2 Y_{20}(\hat{r}) \Big] ,
\label{r3Y10_der}
\\
\nabla_{\pm 1} r^3 Y_{1,\mp 1}(\hat{r}) &=
\frac{1}{\sqrt{3}} \, \Big[{-}5 r^2 Y_{00} + \frac{2}{\sqrt{5}} r^2 Y_{20}(\hat{r})\Big] ,
\label{r3Y11_der}
\end{align}
\begin{equation}
Y_{00} = 1/(2\sqrt{\pi}) ,
\label{Y00}
\end{equation}
\begin{eqnarray}
\label{Y0Y2}
[{\bf Y}^{0}_{1\mu}]^* \cdot {\bf Y}^{2}_{1\mu} &=&
\frac{1}{\sqrt{40\pi}} Y_{20} \, c_{\mu}  ,
\\
\label{Y0Y0}
[{\bf Y}^{0}_{1\mu}]^* \cdot {\bf Y}^{0}_{1\mu} &=& \frac{1}{4\pi}
\end{eqnarray}
where $c_{\mu}$=-2 for $\mu$=0 and 1 for $\mu=\pm 1$.


\begin{thebibliography}{99}
\bibitem{Th61} 
 D.J. Thouless,
 Nucl. Phys. {\bf 22}, 78 (1961).
\bibitem{TV62} 
  D.J. Thouless nd J.G. Valatin,
  Nucl. Phys. {\bf 31}, 211 (1962).
\bibitem{Mar69} 
 E.R. Marshalek and J. Weneser,
  Ann. Phys. (NY) {\bf 53}, 569 (1969).
\bibitem{Row70} 
  D.J. Rowe,
  {\it Nuclear Collective Motion} (Mothuen, London, 1970).
\bibitem{Ri80} 
  P. Ring and P. Schuck,
  {\it Nuclear Many Body Problem} (Springer-Verlag N.Y.-Hedelberg-Berlin, 1980).
\bibitem{BR86} 
   J.P. Blaizot and G. Ripka, {\it Quantum Theory of Finite
   Systems}, Ch. 10 (MIT Press, Cambridge, MA, 1986).
\bibitem{VG81} 
  N. Van Giai and H. Sagawa,
  Nucl. Phys. A {\bf 371}, 1 (1981).
\bibitem{Ha01} 
  M.N. Harakeh, and A. van der Woude,
  {\it Giant Resonances} (Clarendon Press Oxford, 2001).
\bibitem{Ba94} 
  E.B. Balbutsev, I.V. Molodtsova, and A.V. Unzhakova,
   Europhys. Lett. {\bf 26}, 499 (1994).
\bibitem{Kv11}  
  J. Kvasil, V.O. Nesterenko, W. Kleinig, P.G. Reinhard,
  and P. Vesely, Phys. Rev. C {\bf 84}, 034303 (2011).
\bibitem{Re92} 
  P.-G. Reinhard,
  Ann. Phys. (Leipzig) {\bf 504}, 632 (1992).
\bibitem{Be03} 
  M. Bender, P.-H. Heenen, and P.-G. Reinhard,
  Rev. Mod. Phys. {\bf 75}, 121 (2003).
\bibitem{Ni06} 
   T. Nik\v{s}i\'{c}, D. Vretenar, and P. Ring,
   Phys. Rev. C {\bf 74}, 064309 (2006).
\bibitem{RRR18} 
  L.M. Robledo, T.R. Rodr\'iguez, R.R. Rod\'iguez-Guzm\'an,
  arXiv:1807.02518v1[nucl-th].
\bibitem{Co00} 
   G. Colo, N. Van Giai, P.F. Bortignon, and M.R.Qualia,
   Phys. Lett. {\bf{B485}}, 362 (2000).
\bibitem{Ju08} 
   Jun Li, G. Colo, and J. Meng,
   Phys. Rev. C {\bf 78}, 064304 (2008).
\bibitem{Ars10}  
  N.N. Arsenyev abnd A.P. Severyukhin,
  Phys. Part. Nucl. Lett. {\bf 7}, 112 (2010).
\bibitem{Na07}
   T. Nakatsukasa, T. Inakura, and K. Yabana,
   Phys. Rev. C {\bf 76}, 024318 (2007).
\bibitem{Mi12}
  K. Mizuyama  and Gianluca Colo,
   Phys. Rev. C {\bf 85}, 024307 (2012).
\bibitem{Vre00PLB} 
 D. Vretenar, N. Paar, and P. Ring,
 Phys. Lett. {\bf B 487}, 334 (2000).
\bibitem{Vre02tor} 
D. Vretenar, N. Paar, P. Ring, and T. Nik\v{s}i\'{c},
 Phys. Rev. C {\bf 65}, 021301(R) (2002).
\bibitem{Pa07} 
  N. Paar, D. Vretenar, E. Khan, and G. Colo,
  Rep. Prog. Phys. {\bf 70}, 691 (2007).
\bibitem{Kl08} 
 W. Kleinig, V.O. Nesterenko, J. Kvasil, P.-G. Reinhard and P. Vesely,
  Phys. Rev. C {\bf 78}, 044313 (2008).
\bibitem{Kv16E0}  
  J. Kvasil, V.O. Nesterenko, A. Repko, W. Kleinig, and P.-G. Reinhard,
  Phys. Rev. C {\bf 94}, 064302 (2016).
\bibitem{Te10} 
  J. Terasaki and J. Engel,
  Phys. Rev.  C {\bf 82}, 034326 (2010).
\bibitem{Ne16E2} 
 V.O. Nesterenko, V.G. Kartavenko, W. Kleinig,
 J. Kvasil, A. Repko, R.V. Jolos, and P.-G. Reinhard,
 Phys. Rev. C {\bf 93}, 034301 (2016).
\bibitem{KvNa86} 
  J. Kvasil and R.G. Nazmitdinov,
  Sov. J. Part. Nucl. {\bf 17}, 265 (1986).
\bibitem{Na17} 
  H. Nakada, Prog. Theor. Exp. Phys. {\bf 023D02}, 099101 (2016).
\bibitem{Pya77} 
  N.I. Pyatov and D.I. Salamov,
  Nucleonika {\bf 22}, 127 (1977).
\bibitem{Pa67} 
  F. Palumbo, Nuclear Physics {\bf 99}, 100 (1967)
\bibitem{Do05} 
  F. Donau,
  Phys. Rev. Lett. {\bf 94}, 092503 (2005).
\bibitem{Ne02}  
  V.O. Nesterenko, J. Kvasil, and P.-G. Reinhard,
  Phys. Rev. C {\bf 66}, 044307 (2002).
\bibitem{Ne06} 
  V.O. Nesterenko, W. Kleinig, J. Kvasil, P. Vesely, P.-
  G. Reinhard, and D. S. Dolci,
  Phys. Rev. C {\bf 74}, 064306 (2006).
\bibitem{Ser03}
  Ll. Serra, R.G. Nazmitdinov, and A. Puente,
  Phys. Rev. B {\bf 68}, 035341 (2003).
\bibitem{Rep17EPJA} 
 A. Repko, J. Kvasil, V.O. Nesterenko, and P.-G. Reinhard,
 Eur. Phys. J. A {\bf  53}, 221 (2017).
\bibitem{Re15th} 
  A. Repko, {\it Theoretical description of nuclear collective excitations},
  PhD thesis, Math.-Phys. Faculty of Charles
  University in Prague (2015), arXiv:1603.04383 [nucl-th].
\bibitem{Re15ar} 
 A. Repko, J. Kvasil, V.O. Nesterenko, and P.-G. Reinhard,
 arXiv:1510.01248v3 [nucl-th].
\bibitem{Cha98} 
  E. Chabanat, P. Bonche, P. Haensel, J. Meyer, R. Schaeffer,
  Nucl. Phys. {\bf A635}, 231 (1998).
\bibitem{Kv13} 
  J. Kvasil, V.O. Nesterenko, W. Kleinig, D. Bozik, P.-G. Reinhard,
  N. Lo Iudice,
  Eur. Phys. J. {\bf A49}, 119 (2013).
\bibitem{NePRL18} 
  V.O. Nesterenko, A. Repko, J. Kvasil, and P.-G. Reinhard,
Phys. Rev. Lett. {\bf 120}, 182501 (2018).
\bibitem{Ne18ar} 
V.O. Nesterenko, J. Kvasil, A. Repko, and P.-G. Reinhard,
 Eur. Phys. J. Web of Conf. {\bf 194}, 03005 (2018).
\bibitem{Va76} 
  D. A. Varshalovich, A. N. Moskalev, and V. K. Khersonskii,
  {\it Quantum Theory of Angular Momentum} (World Scientific,
  Singapore, 1976).
\bibitem{Gu81} 
  G.M. Gurevich, L.E. Lazareva, V.M. Mazur, S.Yu Merkulov, G.V. Solodukov, and V.A. Tyutin,
  Nucl. Phys. {\bf A351}, 257 (1981).
\bibitem{Yo04} 
  D.H. Youngblood, Y.W. Lui, H.L. Clark, B. John, Y. Tokimoto, and X. Chen,
  Phys. Rev. C{\bf 69}, 034315 (2004).
\bibitem{It03} 
  M. Itoh, H. Sakaguchi,M. Uchida, T. Ishikawa, T. Kawabata, T.
Murakami, H. Takeda, T. Taki, S. Terashima, N. Tsukahara, Y.
Yasuda, M. Yosoi, U. Garg, M. Hedden, B. Kharraja, M. Koss,
B. K. Nayak, S. Zhu, H. Fujimura, M. Fujiwara, K. Hara, H. P.
Yoshida, H. Akimune, M. N. Harakeh, and M. Volkerts,
  Phys. Rev. C{\bf 68}, 064602 (2003).
\bibitem{BM1} 
A. Bohr and B. R. Mottelson, {\it Nuclear Structure} (Benjamin,
New York, 1969), Vol. 1.
\bibitem{Be00} 
 M. Bender, K. Rutz, P.-G. Reinhard, J.A. Maruhn,
 Eur. Phys. J. A {\bf 8}, 59 (2000).
\end{thebibliography}
\end{document}